\documentclass[useAMS,usenatbib]{mn2e}
\usepackage{graphicx}
\usepackage{rotating}

\def \aj {AJ}
\def \mnras {MNRAS}
\def \pasp {PASP}
\def \apj {ApJ}
\def \apjs {ApJS}
\def \apjl {ApJL}
\def \aap {A\&A}
\def \nat {Nature}

\begin{document}

\title[LSQ12btw and LSQ13ccw, two distant Type Ibn SNe]{Massive stars exploding in a He-rich circumstellar medium. VI. Observations of two distant Type Ibn supernova candidates discovered by La Silla-QUEST}

\author[Pastorello et al.]{A. Pastorello,$^1$\thanks{andrea.pastorello@oapd.inaf.it} E. Hadjiyska,$^2$ D. Rabinowitz,$^2$ S. Valenti,$^{3,4}$ M. Turatto,$^1$
\newauthor G. Fasano,$^1$ S. Benitez-Herrera,$^5$ C. Baltay,$^2$ S. Benetti,$^1$ M. T. Botticella,$^6$
\newauthor E. Cappellaro,$^1$  N. Elias-Rosa,$^1$ N. Ellman,$^2$  U. Feindt,$^7$ A. V. Filippenko,$^8$
\newauthor M. Fraser,$^{9}$   A. Gal-Yam,$^{10}$ M. L. Graham,$^8$ D. A. Howell,$^{3,4}$ C. Inserra,$^{11}$
\newauthor P. L. Kelly,$^8$   R. Kotak,$^{11}$   M. Kowalski,$^7$ R. McKinnon,$^2$ A. Morales-Garoffolo,$^{12}$
\newauthor   P. E. Nugent,$^{13,8}$  S. J. Smartt,$^{11}$ K. W. Smith,$^{11}$ M. D. Stritzinger,$^{14}$ M. Sullivan,$^{15}$
\newauthor    S. Taubenberger,$^{5,16}$ E. S. Walker,$^2$ O. Yaron,$^{10}$ and D. R. Young.$^{11}$
\\
$^{1}$INAF-Osservatorio Astronomico di Padova, Vicolo dell'Osservatorio 5,  35122 Padova, Italy\\
$^{2}$Department of Physics, Yale University, P.O. Box 208120, New Haven CT, 06520-8120, USA\\
$^{3}$Las Cumbres Observatory Global Telescope Network, Inc., Santa Barbara, CA 93117, USA\\
$^{4}$Department of Physics, University of California, Santa Barbara, CA 93106-9530, USA\\
$^{5}$Max-Planck-Institut f\"ur Astrophysik, Karl-Schwarzschild-Str. 1, 85741 Garching, Germany\\
$^{6}$INAF -- Osservatorio astronomico di Capodimonte, Salita Moiariello 16, 80131 Napoli, Italy\\
$^{7}$Physikalisches Institut, Universit\"at Bonn, Nu{\ss}allee 12, 53115 Bonn, Germany\\
$^{8}$Department of Astronomy, University of California, Berkeley, CA 94720-3411, USA \\
$^{9}$Institute of Astronomy, University of Cambridge, Madingley Road, Cambridge CB3 0HA, UK \\
$^{10}$Department of Particle Physics and Astrophysics, Faculty of Physics, The Weizmann Institute of Science, Rehovot 76100, Israel\\ 
$^{11}$Astrophysics Research Centre, School of Mathematics and Physics, Queen's University Belfast, Belfast BT7 1NN, UK\\
$^{12}$Institut de Ci\`encies de l'Espai (CSIC-IEEC), Campus UAB,  Torre C5, 2a planta, 08193 Barcelona, Spain\\
$^{13}$Computational Cosmology Center, Computational Research Division, Lawrence Berkeley National Laboratory, 1 Cyclotron Road \\ MS 50B-4206, Berkeley, CA 94720, USA\\
$^{14}$Department of Physics and Astronomy, Aarhus University, Ny Munkegade 120, 8000 Aarhus C, Denmark\\
$^{15}$School of Physics and Astronomy, University of Southampton, Southampton SO17 1BJ, UK\\
$^{16}$European Organisation for Astronomical Research in the Southern Hemisphere (ESO), Karl-Schwarzschild-Str. 2, 85748 Garching \\ bei M\"{u}nchen, Germany
}

\date{Accepted YYYY Month XX. Received YYYY Month XX; in original form YYYY Month XX}

\pagerange{\pageref{firstpage}--\pageref{lastpage}} \pubyear{201X}

\maketitle

\label{firstpage}
\clearpage

\begin{abstract}
We present optical observations of the peculiar stripped-envelope supernovae (SNe) LSQ12btw and LSQ13ccw discovered by the
La Silla-QUEST survey. LSQ12btw reaches an absolute peak magnitude of $M_g = -19.3 \pm 0.2$, and shows an asymmetric
light curve. Stringent prediscovery limits constrain its rise time to
maximum light to less than 4 days, with a slower post-peak luminosity decline, similar to that experienced
by the prototypical SN~Ibn 2006jc.  
LSQ13ccw is somewhat different: while it also exhibits a very fast rise to maximum, it reaches a fainter absolute peak magnitude
($M_g = -18.4 \pm 0.2$), and experiences an extremely rapid post-peak decline similar to that observed in the
peculiar SN~Ib 2002bj. A stringent prediscovery limit and an early marginal detection of LSQ13ccw
allow us to determine the explosion time with an uncertainty of $\pm$1 day.
The spectra of LSQ12btw show the typical narrow He~I emission lines characterising Type Ibn SNe, suggesting
that the SN ejecta are interacting with He-rich circumstellar material. The He~I lines in the spectra of LSQ13ccw exhibit
weak narrow emissions superposed on broad components. An unresolved H$\alpha$ line is also detected, suggesting a 
tentative Type Ibn/IIn classification.
As for other SNe~Ibn, we argue that LSQ12btw and LSQ13ccw likely result from the explosions of 
Wolf-Rayet stars that experienced instability phases prior to core collapse. We inspect the host galaxies 
 of SNe~Ibn, and we show that all of them but one are hosted in spiral galaxies, likely in
environments spanning a wide metallicity range.
\end{abstract}


\section{Introduction} \label{intro}
The discoveries of SN 1999cq \citep{mat00} and, a few years later, the well-studied SN 2006jc \citep[e.g.,][]{pasto07,fol07} revealed the existence of a new
group of core-collapse supernovae (CCSNe) with peculiar observational properties: a relatively high
peak luminosity ($M_R \approx -19$ mag), blue colours, and spectra showing
relatively narrow (a few thousand km s$^{-1}$) emission lines of He~I superposed on typical Type Ic supernova (SN~Ic; e.g., Filippenko 1997) spectral features. 
These observables have been interpreted as signatures of interaction between the SN ejecta and a circumstellar medium (CSM) 
rich in He \citep[][and references therein]{chu09}. Very few SNe 
of this family \citep[named Type Ibn SNe;][]{pasto08a} were discovered in the past,
but their number has significantly grown in recent years, reaching the current number of 16.
These new discoveries are showing that a wide degree of heterogeneity exists in the properties of SNe~Ibn.
A number of papers studying a variety of SNe~Ibn have been published so far,\footnote{Additional publications on this subject include \citet{pasto08a,pasto08b}, \citet{seppo08}, \citet{smi08}, \citet{imm08}, \citet{elisa08}, \citet{noz08}, \citet{tom08}, \citet{anu09}, \citet{sak09}, \citet{smi12}, \citet{san13}, \citet{mod14}, \citet{bia14}, and \citet{gor13}.}
and observations of three recent, unusual SNe~Ibn (SN 2010al, SN 2011hw, and OGLE-2013-SN-006) are published in two companion papers 
of our team \citep{pasto13a,pasto13b}.
It is interesting to note that SN 2011hw, along with another object studied some years ago (SN 2005la), showed  clear evidence of H lines \citep{pasto08b,smi12,pasto13a,mod14,bia14}, indicating that their CSM was also 
moderately H-rich. This led researchers to the conclusion that there is a continuity in properties between SNe~Ibn and SNe~IIn \citep[e.g.,][]{smi12}.

 Because of their rarity, early discovery of new members of this class offers
an opportunity for improving our understanding of their physical nature. 
In this respect, the currently running programs for the prompt spectroscopic classification of
new SN candidates --- such as the Public ESO Spectroscopic Survey for Transient Objects \citep[PESSTO;][]{2013Msngr} and the 
Asiago Classification Program \citep[ACP;][]{2014arXiv1403} -- 
play an important role in providing new SNe~Ibn as well as other types of unusual transients. 

This paper describes our dataset for two new peculiar stripped-envelope SNe that we propose to be
SNe~Ibn, and discusses them in the context of what is currently known on the subject. 


\begin{figure}
{\includegraphics[width=3.4in]{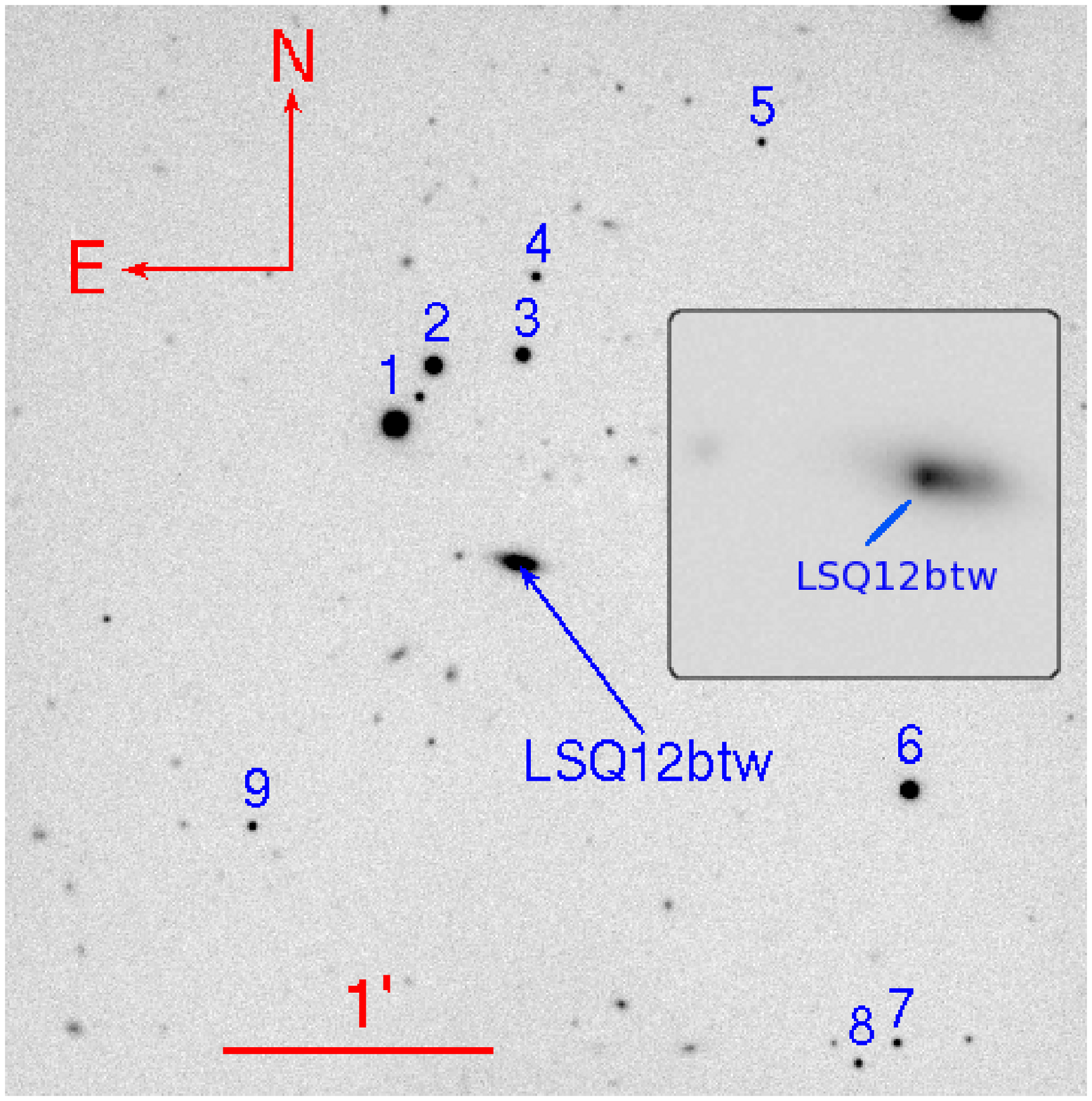}
\includegraphics[width=3.4in]{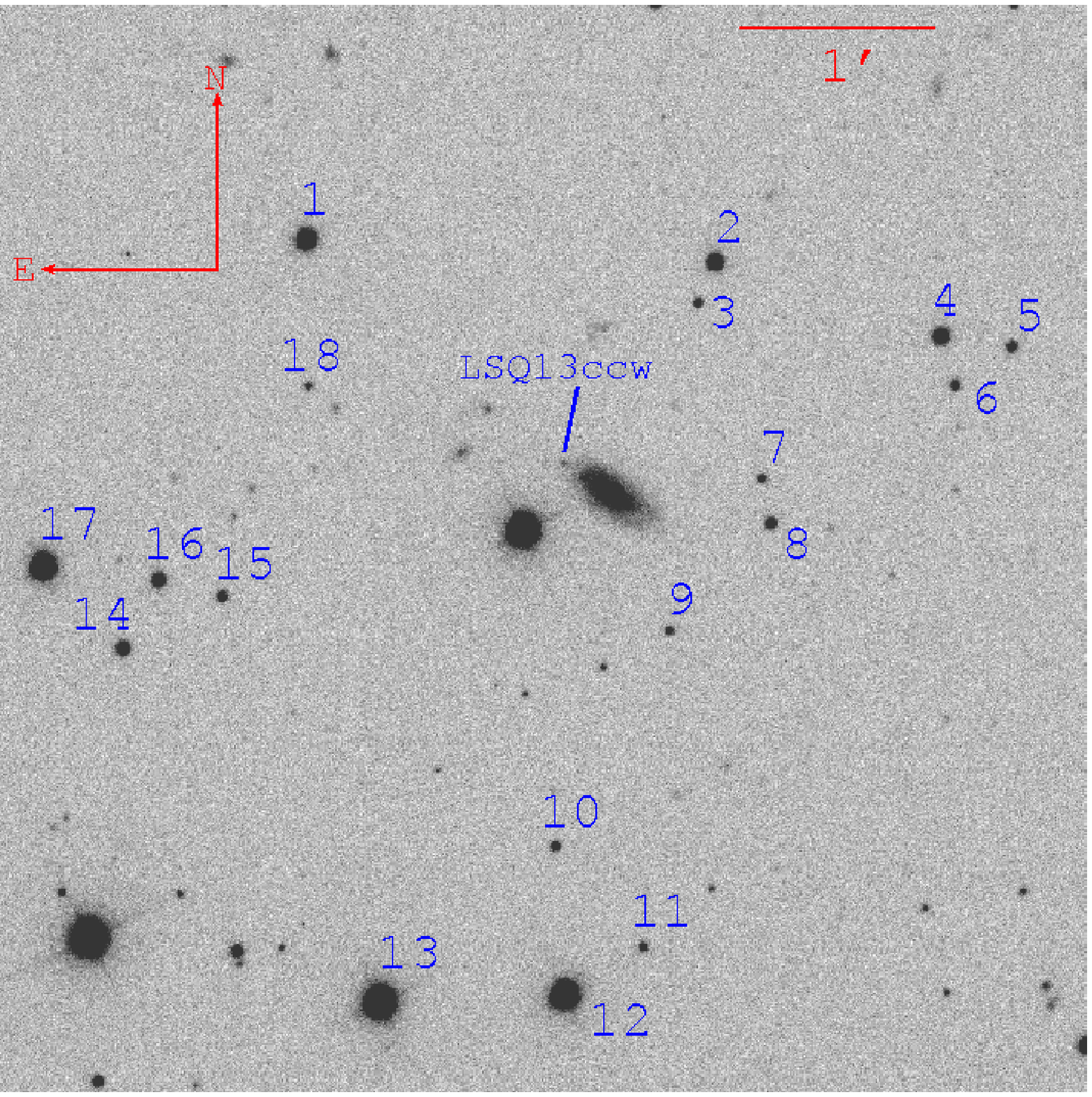}}
\caption{Top: ESO-NTT $r$-band image of LSQ12btw obtained on 2012 March 14. The insert at right shows the nuclear region 
of the host galaxy with a different cut level, highlighting the presence of the SN, due east of the galaxy nucleus. 
Bottom: Liverpool Telescope $V$-band image of LSQ13ccw and its host galaxy obtained on 2013 September 22. 
The numbers mark the SDSS reference stars used to calibrate the magnitudes of the two SNe.} 
\label{fig1}
\end{figure}

LSQ12btw and LSQ13ccw were found by the La Silla-QUEST (LSQ) survey \citep{had11,bal13} and tentatively
classified as SN 2006jc-like (Type Ibn) SNe by the PESSTO collaboration
\citep{val12,kan13}.
The former object was discovered on 2012 April 9.04 (UT dates are used
throughout this paper), close to the nucleus of the spiral, star-forming galaxy SDSS J101028.69+053212.5 (see Fig. \ref{fig1}). 
SDSS DR3 reports a redshift $z = 0.05764 \pm 0.00006$.\footnote{The spectroscopic data were obtained on 2004 September 28; http://www.sdss.org/dr3/products/spectra/getspectra.html.}
Adopting a Hubble constant H$_0$ = 73 km s$^{-1}$ Mpc$^{-1}$ ($\Omega_{\rm M}$ = 0.27 and $\Omega_\Lambda$ = 0.73),
we infer a luminosity distance of about 247 $\pm$ 16 Mpc ($\mu$ = 36.97 $\pm$ 0.15 mag). 
No sign of  interstellar absorption within the host galaxy is visible in the SN spectra (see Sec.~\ref{s_sp}), so we consider only the Galactic extinction  
component for LSQ12btw, $A_g = 0.075$ mag \citep{sch11}.  

LSQ13ccw was discovered on 2013 September 4.03 near the galaxy 2MASX J21355077-1832599 (also known as PGC 866755; Fig. \ref{fig1}).
The redshift of the host galaxy was measured through the SN features and the detection of weak and narrow 
lines of [O~III] $\lambda\lambda$4959, 5007 from the galaxy. These give $z = 0.0603 \pm 0.0002$ and a luminosity distance of 259 $\pm$ 17 Mpc ($\mu$ = 37.07 $\pm$ 0.14 mag).
Since there is no clear spectroscopic evidence (see Sec. \ref{s_sp}) of host-galaxy absorption, we again adopt only the Milky Way component
($A_g = 0.163$ mag) as total line-of-sight absorption to LSQ13ccw.

\section{Observations}  \label{obs}

\subsection{Photometry} \label{s_pho}

The field of LSQ12btw was repeatedly monitored over the past few years by LSQ and Pan-STARRS1 (PS1) \citep[see][for a description of the PS1 3Pi survey, exposure times, and depths]{ins13}.
As we will discuss later in this section, there is no previous detection of variability at the SN position. 
The most recent LSQ image with no detection was obtained on 2012 March 23.12 (JD = 2,456,009.62), while images on March 27.11 (JD = 2,456,013.61) clearly showed the new source. This allows us to constrain
the epoch of shock breakout to JD = 2,456,010.0$^{+3.6}_{-1.0}$.

\begin{figure*}
{\includegraphics[width=3.45in]{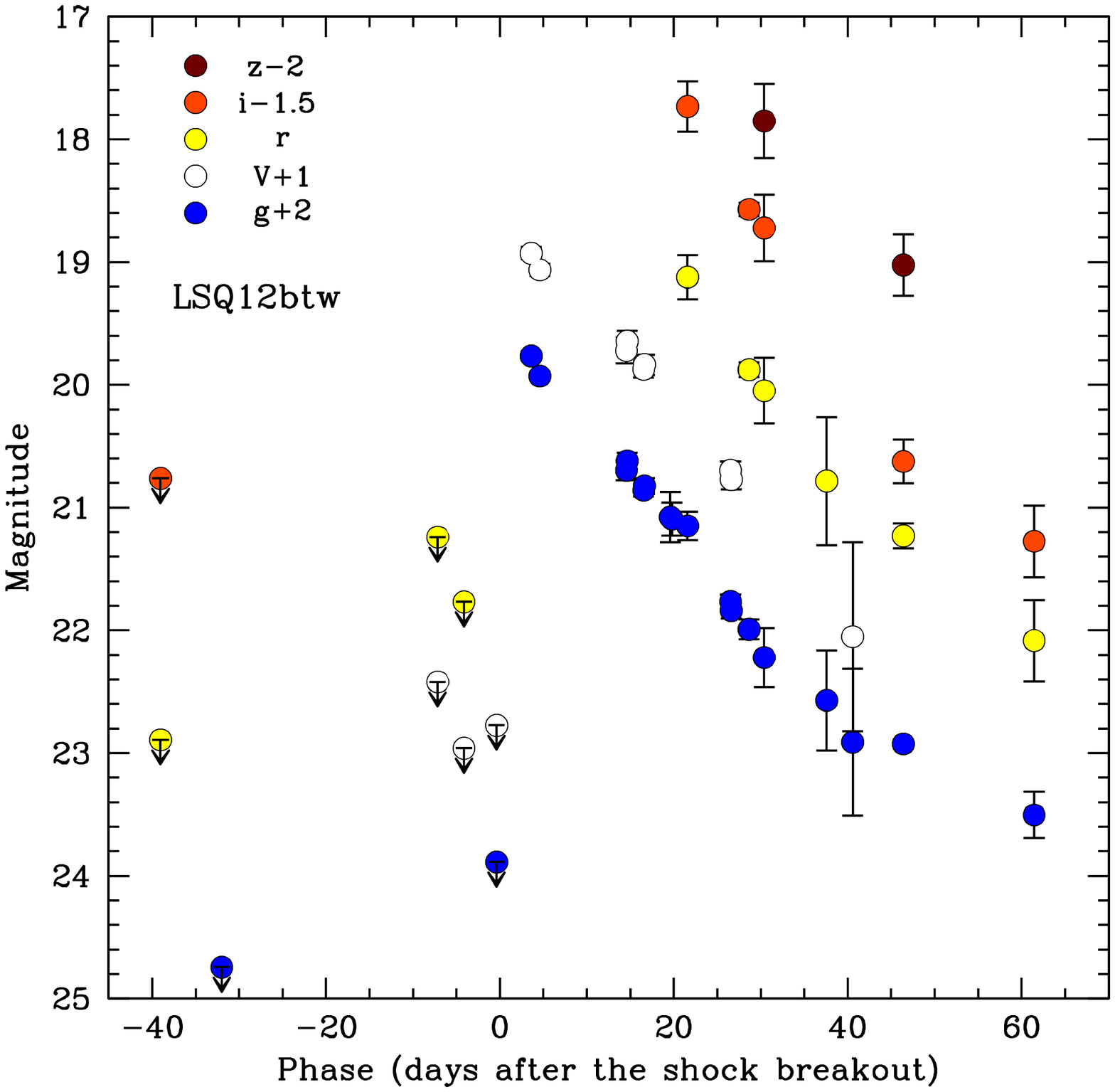}
\includegraphics[width=3.45in]{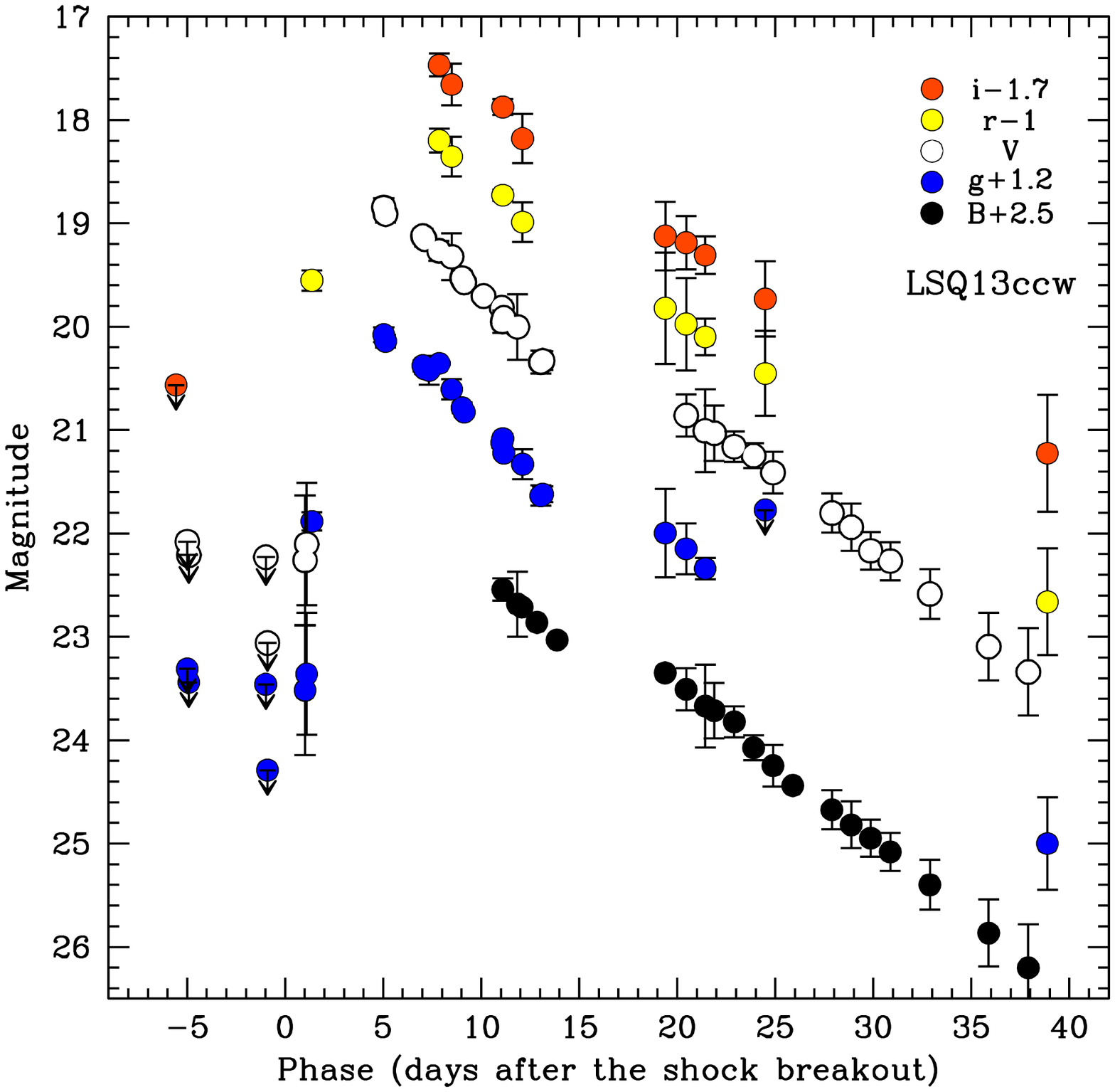}}
\caption{Left: Optical light curves of LSQ12btw. Right: Optical light curves of LSQ13ccw. 
$B$ and $V$ light curves were calibrated using the \protect\citet{lan92} catalogue (Vegamag system), while $g$, $r$, $i$, and $z$ light
curves were calibrated using the Sloan catalogue  (ABmag system).} 
\label{fig2}
\end{figure*}

\begin{table*}  
\begin{center}
\caption{Sloan $g$, $r$, $i$, and $z$ photometry of LSQ12btw, and associated uncertainties.
\label{tab:1}
}
\begin{tabular}{ccccccccc}
\hline \hline
Date & JD & Phase$^\ddag$ & $V$ & $g$ & $r$ & $i$ & $z$ & Inst. \\
 \hline
13/02/2012 & 2455970.97  & $-$39.0 & & --           & $>$22.89     & $>$22.26     &   --    &   1  \\
20/02/2012 & 2455978.02  & $-$32.0 & & $>$22.74     &  --          &  --          &   --    &   1  \\
16/03/2012$^{\dag}$ & 2456002.84  & $-$7.2 & [$>$21.42] & --     & $>$21.24     & --     &   --    &  1  \\
19/03/2012$^{\dag}$ & 2456005.87  & $-$4.1 & [$>$21.96] & --     & $>$21.77     & --     &   --    &  1  \\
23/03/2012$^{\ast}$ & 2456009.62  & $-$0.4 & [$>$21.77] & $>$21.89     & --     &   --   &   --    &  2 \\
27/03/2012$^{\ast}$ & 2456013.61  & +3.6 &  [17.93 0.05] & 17.76 0.04 & -- &   --         &   --    &   2 \\
28/03/2012$^{\ast}$ & 2456014.61  & +4.6 &  [18.06 0.05] & 17.93 0.04 & -- &   --         &   --    &   2 \\
07/04/2012$^{\ast}$ & 2456024.54  & +14.5 & [18.72 0.11] & 18.70 0.08 & -- &   --         &   --    &   2 \\ 
07/04/2012$^{\ast}$ & 2456024.63  & +14.6 & [18.64 0.09] & 18.62 0.07 & -- &   --         &   --    &   2 \\
09/04/2012$^{\ast}$ & 2456026.54  & +16.5 & [18.87 0.07] & 18.86 0.05 & -- &   --         &   --    &   2 \\
09/04/2012$^{\ast}$ & 2456026.62  & +16.6 & [18.84 0.08] & 18.82 0.06 & -- &   --         &   --    &   2 \\
12/04/2012 & 2456029.59  & +19.6 & & 19.08 0.20 &   --         &   --         &   --    &    3 \\
12/04/2012 & 2456029.80  & +19.8 & & 19.09 0.13 &   --         &   --         &   --    &    1 \\
14/04/2012 & 2456031.58  & +21.6 & & 19.15 0.12 & 19.12 0.18 & 19.23 0.21 &   --    &    3 \\
19/04/2012$^{\ast}$ & 2456036.52  & +26.5 & [19.70 0.07] & 19.77 0.06 & -- &   --         &   --    &   2 \\
19/04/2012$^{\ast}$ & 2456036.61  & +26.6 & [19.77 0.08] & 19.84 0.06 & -- &   --         &   --    &   2 \\
21/04/2012 & 2456038.67  & +28.7 & & 19.99 0.08 & 19.88 0.06 & 20.07 0.06 &   --    &    3 \\
22/04/2012 & 2456040.39  & +30.4 & & 20.22 0.24 & 20.05 0.27 & 20.22 0.27 & 19.85 0.30 &  4 \\
30/04/2012 & 2456047.58  & +37.5 & & 20.57 0.41 & 20.79 0.52 &   --         &   --    &   3  \\
03/05/2012$^{\ast}$ & 2456050.57  & +40.6 & [21.05 0.77] & 20.91 0.60 & -- &   --         &   --    &   2 \\ 
08/05/2012 & 2456056.40  & +46.4 & & 20.93 0.04 & 21.23 0.10 & 22.12 0.18 & 21.02 0.25 &  5  \\
23/05/2012 & 2456071.45  & +61.5 & & 21.50 0.19 & 22.08 0.33 & 22.77 0.29 &   --    &  6 \\ \hline
\end{tabular}

\begin{flushleft}
Notes: Measurements from LSQ and PS1 wide
broadband images were first scaled to Sloan $g$ and Sloan $r$ magnitudes, respectively, and then transformed to 
Johnson $V$-band magnitudes by applying the relations of \protect\citet{jes05} for blue stars (see text).
When no source was detected at the SN position, 3$\sigma$ detection limits are reported. PS1 photometry was obtained 
from the combined images after coadding two single exposures. Indicated errors account for both the uncertainties 
in the photometric measurement and the final zeropoint calibration.\\

1 = 1.8-m Pan-STARRS1 (PS1) Telescope + GPC1;
2 = 1.0-m ESO Schmidt Telescope + La Silla QUEST Camera;
3 = 3.58-m ESO New Technology Telescope + EFOSC2;
4 = 2.0-m Liverpool Telescope + RATCam;
5 = 3.58-m Telescopio Nazionale Galileo + Dolores;
6 = 10.4-m Gran Telescopio Canarias (GranTeCan) + Osiris.\\
$^{\ddag}$Days from the time of the shock breakout;
$^{\dag}$Pan-STARRS1 $w_{\rm P1}$-filter data converted to Sloan $r$ and Johnson $V$ magnitudes;\\
$^{\ast}$LSQ $Q_{\rm st}$-filter data  converted to Sloan $g$  and Johnson $V$ magnitudes.
\end{flushleft}

\end{center}
\end{table*}

\begin{table*}  
\begin{center}
\caption{Sloan $g$, $r$, $i$, and $z$ photometry, and Johnson-Bessell $B$ and $V$ photometry, of LSQ13ccw, and associated uncertainties.
\label{tab:2}
}
\begin{tabular}{ccccccccc}
\hline \hline
Date & JD & Phase$^\ddag$ & $B$ & $V$ & $g$ & $r$ & $i$ & Inst. \\
 \hline
24/08/2013         &  2456528.94 & $-$5.6 & -- &  -- & -- & -- & $>$22.26 & 1 \\
25/08/2013$^{\ast}$ &  2456529.51 & $-$5.0 & -- &  [$>$22.08] & $>$22.11 & -- & -- & 2 \\
25/08/2013$^{\ast}$ &  2456529.59 & $-$4.9 & -- &  [$>$22.21] & $>$22.24 & -- & -- & 2 \\
29/08/2013$^{\ast}$ &  2456533.52 & $-$1.0 & -- &  [$>$22.23] & $>$22.26 & -- & -- & 2 \\
29/08/2013$^{\ast}$ &  2456533.60 & $-$0.9 & -- &  [$>$23.06] & $>$23.09 & -- & -- & 2 \\
31/08/2013$^{\ast,\P}$& 2456535.52 & +1.0 & -- &  [22.26 0.63] & 22.32 0.63 & -- & -- & 2 \\
31/08/2013$^{\ast,\P}$& 2456535.60 & +1.1 & -- &  [22.10 0.59] & 22.16 0.59 & -- & -- & 2 \\
31/08/2013         &  2456535.87 & +1.4 & -- &  -- & 20.68 0.09 & 20.55 0.10 & -- & 1 \\
04/09/2013$^{\ast}$ &  2456539.53 & +5.0 & -- &  [18.85 0.09] & 18.88 0.07 & -- & -- & 2 \\
04/09/2013$^{\ast}$ &  2456539.63 & +5.1 & -- &  [18.91 0.08] & 18.94 0.06 & -- & -- & 2 \\
06/09/2013$^{\ast}$ &  2456541.53 & +7.0 & -- &  [19.12 0.07] & 19.18 0.05 & -- & -- & 2 \\
06/09/2013$^{\ast}$ &  2456541.62 & +7.1 & -- &  [19.15 0.07] & 19.21 0.05 & -- & -- & 2 \\
06/09/2013         &  2456541.85 & +7.4 & -- &  -- & 19.22 0.14 & -- & -- & 1 \\ 
06/09/2013         &  2456542.35 & +7.9 & 19.37 0.06 & 19.27 0.10 & 19.16 0.04 & 19.20 0.12 & 19.17 0.11 & 3 \\
07/09/2013         &  2456542.99 & +8.5 & 19.45 0.18 & 19.32 0.23 & 19.41 0.10 & 19.35 0.19 & 19.36 0.20 & 4 \\
08/09/2013$^{\ast}$ &  2456543.53 & +9.0 & -- &  [19.53 0.07] & 19.59 0.05 & -- & -- & 2 \\
08/09/2013$^{\ast}$ &  2456543.64 & +9.1 & -- &  [19.57 0.07] & 19.63 0.04 & -- & -- & 2 \\
09/09/2013         &  2456544.62 & +10.1 & -- &  19.70 0.06 &  -- & -- & -- & 5\\
10/09/2013$^{\ast}$ &  2456545.54 & +11.0 & -- &  [19.82 0.07] & 19.92 0.05 & -- & -- & 2 \\
10/09/2013 &  2456545.60 & +11.1 & 20.04 0.13 & 19.95 0.11 & 19.88 0.04 & 19.73 0.06 & 19.58 0.08 & 6 \\
10/09/2013$^{\ast}$ &  2456545.65 & +11.2 & -- &  [19.91 0.06] & 20.02 0.04 & -- & -- & 2 \\
10/09/2013 &  2456546.35 & +11.9 & 20.18 0.20 & 20.00 0.32 & -- & -- & -- & 3 \\
11/09/2013 &  2456546.60 & +12.1 & 20.21 0.37 & -- & 20.13 0.14 & 19.99 0.19 & 19.88 0.24 & 7 \\
11/09/2013 &  2456547.35 & +12.9 & 20.36 0.25 & -- & -- & -- & -- & 3 \\
12/09/2013$^{\ast}$ &  2456547.54 & +13.0 & -- &  [20.34 0.11] & 20.43 0.10 & -- & -- & 2 \\
12/09/2013$^{\ast}$ &  2456547.64 & +13.1 & -- &  [20.33 0.09] & 20.42 0.08 & -- & -- & 2 \\
12/09/2013 &  2456548.37 & +13.9 & 20.53 0.16 & -- & -- & -- & -- & 8 \\
18/09/2013 &  2456553.88 & +19.4 & 20.85 0.40 & -- & 20.80 0.43 & 20.82 0.54 & 20.82 0.33 & 9 \\
19/09/2013 &  2456554.96 & +20.5 & 21.01 0.21 & 20.86 0.20 & 20.95 0.25 & 20.98 0.45 & 20.89 0.26 & 9 \\
20/09/2013 &  2456555.93 & +21.4 & 21.17 0.17 & 21.01 0.40 & 21.14 0.10 & 21.10 0.18 & 21.01 0.18 & 9 \\
20/09/2013 &  2456556.39 & +21.9 & 21.21 0.32 & 21.03 0.27 & -- & -- & -- & 10 \\
21/09/2013 &  2456557.41 & +22.9 & 21.32 0.23 & 21.16 0.15 & -- & -- & -- & 10 \\
22/09/2013 &  2456558.40 & +23.9 & 21.57 0.10 & 21.25 0.12 & -- & -- & -- & 10 \\
23/09/2013 &  2456558.99 & +24.5 & -- & -- & $>$20.57 & 21.45 0.41 & 21.43 0.36 & 9 \\
23/09/2013 &  2456559.39 & +24.9 & 21.75 0.10 & 21.41 0.20 & -- & -- & -- & 10 \\
24/09/2013 &  2456560.40 & +25.9 & 21.94 0.15 & -- & -- & -- & -- & 10 \\
26/09/2013 &  2456562.40 & +27.9 & 22.17 0.21 & 21.80 0.19 & -- & -- & -- & 10 \\
27/09/2013 &  2456563.38 & +28.9 & 22.32 0.11 & 21.94 0.23 & -- & -- & -- & 10 \\
28/09/2013 &  2456564.37 & +29.9 & 22.45 0.18 & 22.17 0.18 & -- & -- & -- & 10 \\
29/09/2013 &  2456565.37 & +30.9 & 22.58 0.09 & 22.27 0.18 & -- & -- & -- & 10 \\
01/10/2013 &  2456567.39 & +32.9 & 22.90 0.19 & 22.59 0.24 & -- & -- & -- & 10 \\
04/10/2013 &  2456570.39 & +35.9 & 23.37 0.43 & 23.09 0.33 & -- & -- & -- & 10 \\
06/10/2013 &  2456572.41 & +37.9 & 23.70 0.24 & 23.34 0.42 & -- & -- & -- & 10 \\
07/10/2013 &  2456573.38 & +38.9 & -- & -- & 23.80 0.45 & 23.66 0.52 & 22.92 0.57 & 10 \\ \hline
\end{tabular}

\begin{flushleft}

Notes: Measurements from LSQ $Q_{\rm st}$ 
wide broadband images were first scaled to Sloan $g$ magnitudes, and then transformed to 
Johnson $V$-band magnitudes by applying the relations of  \protect\citet{jes05} for blue stars (see text). When 
no source is detected at the SN position, 3$\sigma$ detection limits are reported. Indicated errors account for both the uncertainties 
in the photometric measurement and the final zeropoint calibration.\\

1 = 1.8-m Pan-STARRS1 (PS1) Telescope + GPC1;
2 = 1.0-m ESO Schmidt Telescope + La Silla QUEST Camera;
3 = LCOGT 1m0-13 (SAAO); 
4 = LCOGT 1m0-03 (SSO);
5 = 3.58-m NTT + EFOSC2
6 = LCOGT 1m0-09 (CTIO);
7 = LCOGT 1m0-04 (CTIO); 
8 = LCOGT 1m0-12 (SAAO);
9 = 2.0-m Faulkes Telescope South + fs03;
10 = 2.0-m Liverpool Telescope + I0:0. \\

$^{\ddag}$Days from the time of the shock breakout; $^{\P}$2$\sigma$ detection;\\
$^{\ast}$LSQ $Q_{\rm st}$-filter photometry is converted to Sloan $g$ and Johnson $V$ magnitudes.

\end{flushleft}

\end{center}
\end{table*}

Even more stringent is the constraint on  the explosion epoch for LSQ13ccw. The object was not detected   
on 2013 August 29.10 (JD = 2,456,533.60), while there are two marginal detections on 2013 August 31.02 and 31.10
(JD = 2,456,535.52 and 2,456,535.60). Two additional images were obtained on September 4, showing the object at 
peak luminosity. The fast rise to maximum and the marginal detections of August 31 allow us to 
constrain the time of shock breakout with a small uncertainty, JD = 2,456,534.5 $\pm$ 1.0.

\begin{figure*}
\includegraphics[scale=.52,angle=270]{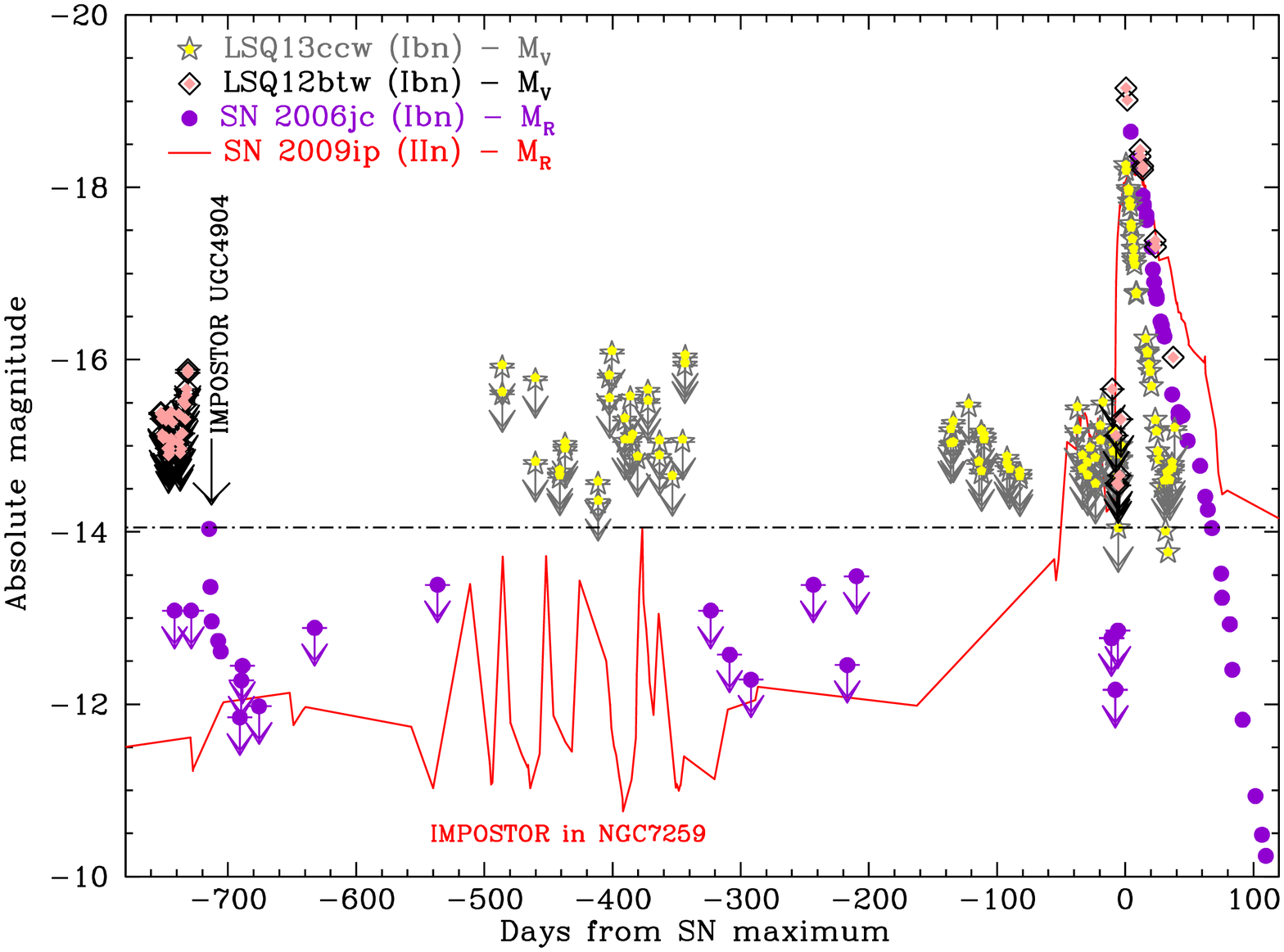}
\caption{Absolute $V$-band light curves of LSQ12btw and LSQ13ccw, compared with the $R$-band observations of the Type Ibn SN 2006jc and the Type IIn SN 2009ip. These observations include
detection limits, pre-SN outbursts, and SN light curves. Data for SN 2006jc are from \protect\citet{pasto07,pasto08a}, while those for SN 2009ip are from \citet{pasto13}, \citet{fra13b}, \citet{mar14}, and \citet{gra14}.
The horizontal dot-dashed line indicates the peak magnitudes of typical SN impostors \citep[e.g.,][]{van00,mau06,tar14}.}
\label{fig3}
\end{figure*}

Photometric data for LSQ12btw and LSQ13ccw were obtained with Sloan filters, using different instruments available to our collaboration 
(see footnotes of Tables \ref{tab:1} and \ref{tab:2} for details).
The data were reduced following standard prescriptions in the IRAF\footnote{IRAF is distributed by the National Optical Astronomy Observatory, which is operated by the Association of 
Universities for Research in Astronomy (AURA), Inc., under cooperative agreement with the US National Science Foundation (NSF).} environment (including bias, overscan, flat-field correction, and final trimming).
Additional Johnson $B$- and $V$-band observations were obtained for LSQ13ccw using LCOGT facilities and the Liverpool Telescope. Standard fields from the \citet{lan92} and \citet{smi02}
catalogues were observed during a few photometric nights, and were used to calibrate the magnitudes of
several stars (used as local standards; Figure \ref{fig1}) in the fields of the two SNe. 
The magnitudes of each of the two transients were calibrated relative to the average magnitudes of the respective sequences of local standards.

Most PS1 and LSQ observations were obtained through very broad filters.\footnote{The broad interference filter used by the LSQ survey,  
labeled as $Q_{\rm st}$, has an almost constant transmission between 460 and 700 nm \citep[see Figure 2 of][]{bal13}, 
while the $w_{\rm P1}$ PS1 filter has a flat transmission between $\sim410$ and 805 nm \citep{ton12}.} 
The magnitudes in the original photometric systems of the two telescopes were scaled to Sloan $g$ (LSQ) and Sloan $r$ (PS1) using the magnitudes of 
Sloan reference stars and the transmission-curve information of the two nonstandard wide passband filters.
Finally, Sloan $g$ or $r$ magnitudes were converted to Johnson $V$ by applying the transformation relations of  \protect\citet{jes05} for blue stars, 
and using the $g-r$ colour information inferred from photometry obtained on adjacent nights and from the available spectra.
The final magnitudes of LSQ12btw and LSQ13ccw are reported in Tables \ref{tab:1} and \ref{tab:2}, respectively, and their multi-band light curves are 
shown in Figure \ref{fig2}. 

\begin{figure*}
\includegraphics[scale=.65,angle=270]{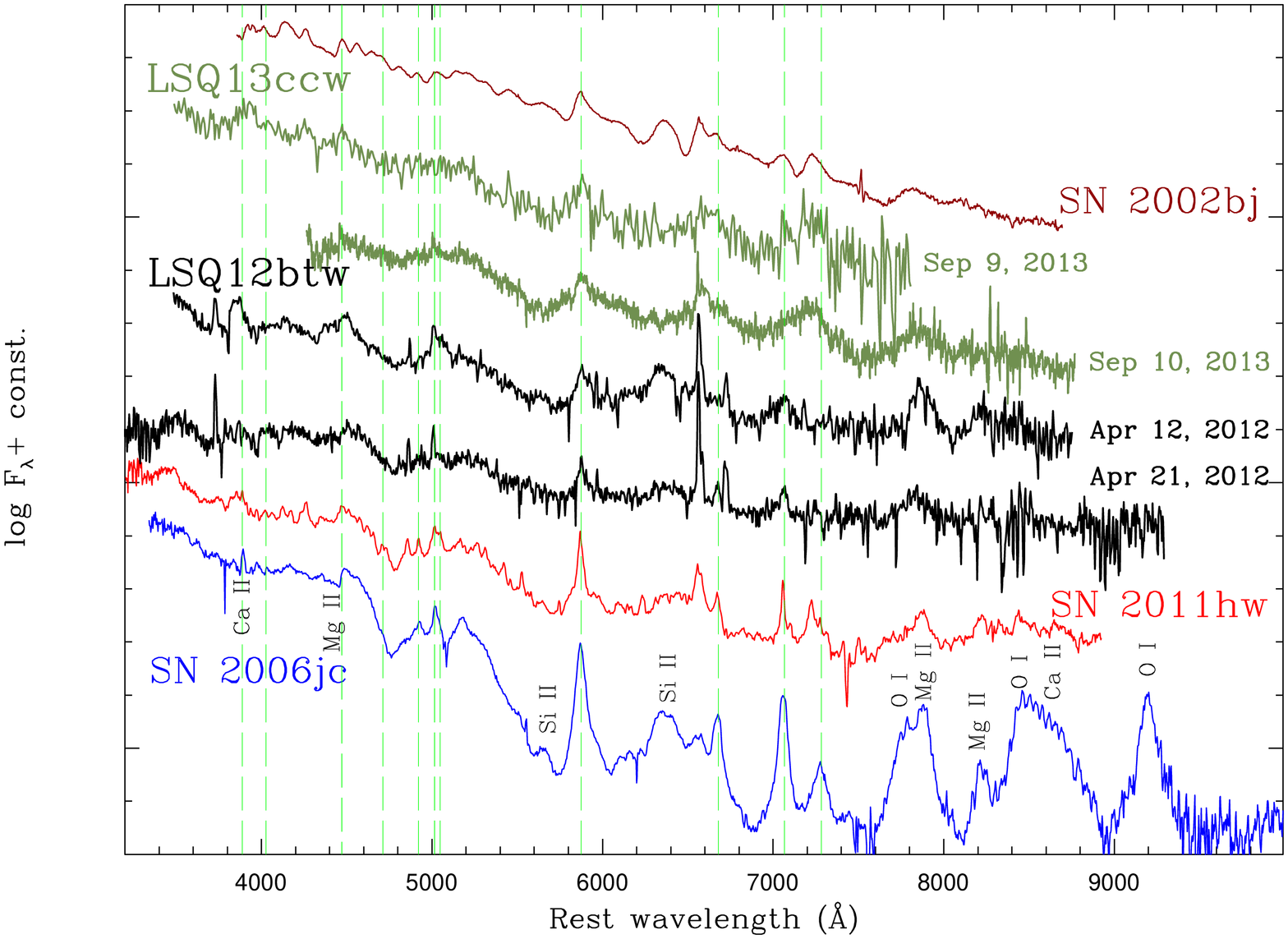}
\caption{Post-peak spectra of LSQ12btw and LSQ13ccw are compared with those of the Type Ibn SNe 2011hw \citep{pasto13a} and 2006jc \citep{pasto07}, and of the peculiar SN 2002bj
\protect\citep[]{poz10}. The spectra of SNe 2011hw and 2006jc were obtained 24 and 25 days after explosion, respectively, while that of SN 2002bj was obtained
7 days after its discovery, $< 2$ weeks after the SN explosion. 
The vertical dashed lines mark the positions of the strongest He~I features. The identification of other lines mentioned in the text has
been done following \citet{pasto07} and \citet{anu09}.
All spectra were corrected for redshift, bringing them to the rest frame. 
The low signal-to-noise ratio 2013 September 9 spectrum of LSQ13ccw has been rebinned to 9~\AA\ per bin.
} 
\label{fig4}
\end{figure*}

\begin{table*}
\begin{center}
\caption{Log of spectroscopic observations of LSQ12btw and LSQ13ccw. 
\label{tab_sp}}
\begin{tabular}{cccccccc}
\hline \hline
Object & Date & JD & Phase & Instrument configuration & Range (\AA) & Resolution (\AA) \\ \hline
LSQ12btw & 12/04/2012 & 2456029.60 & +19.6 & NTT + EFOSC2 + gr.13 & 3650--9250 & 18 \\
LSQ12btw & 21/04/2012 & 2456038.61 & +28.6 & NTT + EFOSC2 + gr.11 + gr.16 & 3350--10,000 & 13,13 \\ \hline
LSQ13ccw & 09/09/2013 & 2456544.63 & +10.1 & NTT + EFOSC2 + gr.13 & 3650--9250 & 18 \\
LSQ13ccw & 10/09/2013 & 2456545.90 & +11.4 & Keck II + DEIMOS + gt.600 l/mm & 4450--9650 & 3.7 \\
 \hline
\end{tabular}

Note: Phases are in days from shock breakout. 
The resolution is measured from the FWHM of the night-sky lines. 
\end{center}
\end{table*}

The light curves of  LSQ12btw are asymmetric, with a fast ($\leq 4$ days) rise to maximum light, and a slower post-peak decline. 
Nonetheless, in analogy to SN 2006jc \citep{pasto08a}, the decline of the light curve of  LSQ12btw is much steeper 
(on average, $\gamma_V = 8.4 \pm 0.3$ mag/100~d) than those observed in canonical SNe~Ib/c. 
We also note that, at phases $>30$ days, the decline slightly slows down, possibly marking the transition to the nebular phase.
The light curve of LSQ13ccw shows a very rapid rise to maximum light similar to that of LSQ12btw, and a surprisingly fast 
average decline $\gamma_V = 13.3 \pm 0.2$ mag/100~d. The light curve of LSQ13ccw is remarkably similar to that of the peculiar SN~Ib 2002bj 
\citep[][cf. Section \ref{disc}]{poz10}. With the distances and reddening values adopted in Section \ref{intro}, the absolute peak magnitudes
of LSQ12btw and LSQ13ccw are $M_g = -19.3 \pm 0.2$ and  $M_g = -18.4 \pm 0.2$, respectively.

One of the most compelling observational results in recent years is the discovery of outbursts occasionally occurring a short time (weeks to years) before
the explosion of interacting SNe \citep[e.g.,][]{pasto07,mau13,ofek13,fra13}. We searched for evidence of pre-SN activity in archival images of both LSQ12btw and LSQ13ccw.
We inspected deep images of the field of LSQ12btw obtained by the LSQ survey (limiting magnitude of about 22), and
by PS1 (limiting magnitude of $\sim$20) in 2010 March. Deep PS1 and LSQ images were also obtained in 2012 March. All these archive images showed 
no source detected at the SN position. PS1 observed the position of LSQ13ccw on 19 occasions (when the object fell on good pixels) between
June 2009 and seven days before our first discovery epoch in the various $grizy$ filters of PS1. LSQ images were collected
from 2012 May to 2012 September, and from 2013 April to August. No pre-SN eruption of LSQ13ccw  was detected.  Although PS1 and 
LSQ have made serendipitous discoveries \citep[see, e.g.,][]{fra13}, the lack of detections at the position of LSQ13ccw and LSQ12btw does not rule out
that such outbursts occurred, since the temporal coverage is sparse and the detection limits are not deep enough
to detect intrinsically faint transients. This can be visualised by comparing the pre-discovery detection limits of the two SNe
with observations of the fields of two well-studied objects showing signs of pre-SN activity: the Type Ibn SN 2006jc in UGC 4904 and
the Type IIn SN 2009ip in NGC 7259 (Figure \ref{fig3}). The pre-SN outbursts of SN 2006jc and 2009ip reached an absolute magnitude of
about $-14$, which is below the detection threshold of almost all our observations of LSQ12btw and LSQ13ccw. 

\subsection{Spectroscopy} \label{s_sp}

The faint apparent magnitudes coupled with the rapid evolution did not allow extensive spectroscopic coverage of these two SNe.
We obtained two spectra of LSQ12btw and one of LSQ13ccw with the 3.58-m New Technology Telescope (NTT) of the European Southern Observatory 
(ESO, La Silla, Chile) equipped with EFOSC2, plus an additional spectrum of LSQ13ccw with the 10-m Keck-II telescope equipped with the DEIMOS 
spectrograph. Spectra of the two SNe and the flux standard stars were obtained with the slit  oriented along the parallactic angle \citep{fil82}.
Information on the spectroscopic observations is reported in Table \ref{tab_sp}.
All spectra\footnote{These spectra are available from WISeREP \protect\citep{yar13}.} are shown in Figure \ref{fig4},
along with those of the Type Ibn SN 2006jc \citep{pasto07}, the transitional SN Ibn/IIn 2011hw \citep{pasto13a}, and a spectrum of the peculiar Type Ib SN 2002bj \citep{poz10}.

\begin{figure}
\includegraphics[scale=.43,angle=0]{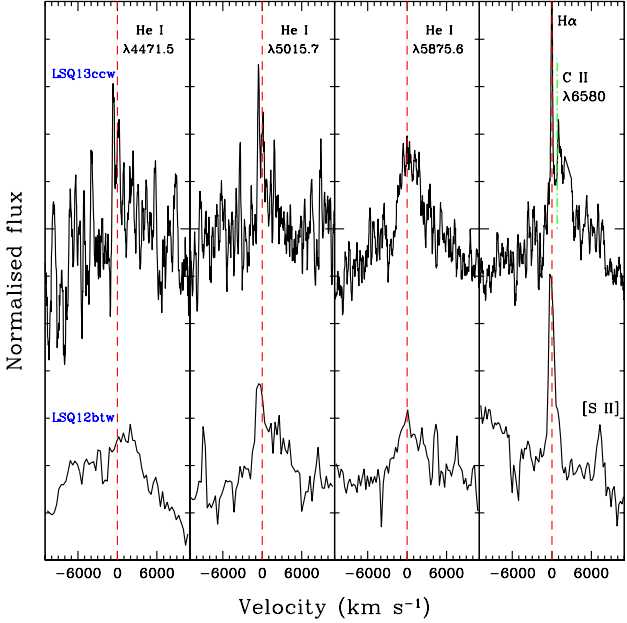}
\caption{Profiles of three He~I lines ($\lambda$4471, $\lambda$5016, $\lambda$5876) and the H$\alpha$ region in the Keck-II spectrum 
of LSQ13ccw obtained on 2013 September 9, and in the NTT spectrum of LSQ12btw obtained on 2012 April 12. Dashed red vertical lines in the 
four panels mark the zero-velocity position of the He~I and H$\alpha$ lines; the dot-dashed green line in the right panel is centered at the
average position of the C~II $\lambda\lambda$6578, 6583 doublet.}
\label{fig5}
\end{figure}

The first spectrum of LSQ12btw, obtained
on 2012 April 12 ($\sim 20$ days after the shock breakout), is the classification spectrum announced by \citet[][]{val12}.
The second spectrum was collected on April 21 ($\sim +29$ days). Between the two epochs there was little evolution, 
with the latter spectrum being slightly redder than
the former. Both show a blue pseudo-continuum, with a  drop in the flux above 5700~\AA. This pseudo-continuum is a common feature in Type Ibn 
and other interacting SNe, and has been attributed to blends of Fe~II emission lines formed in the shocked material \citep{smi09,smi12}.
The spectrum shows many of the lines  observed in SNe~IIn (Fig. \ref{fig4}),\footnote{The comprehensive line identification,
based on the features observed in the higher-quality spectrum of SN 2006jc, is shown in Fig. \ref{fig4} \citep{pasto07,anu09}.} including the broad bump at 6300--6400~\AA\ (possibly due to a blend of
Si~II $\lambda$6355, Mg~II $\lambda$6346, and/or (less likely) [O~I] 
$\lambda\lambda$6300, 6364). 
Another feature at about 7800~\AA\ is clearly detected, probably a line blend including
O~I $\lambda$7774. The near-infrared (NIR) Ca~II triplet,  prominent in SN 2006jc \citep{pasto07}, is weak or absent.
But the most remarkable features are the emissions lines of He~I, with a velocity of 4000--4500 km s$^{-1}$, as measured from the
full width at half-maximum intensity (FWHM) of the He~I $\lambda$5876 line.\footnote{Although we admit that lines with $v_{\rm FWHM} \approx 4000$--4500 km s$^{-1}$
cannot be formally considered to be ``narrow,'' here we consider these velocities as lying in the range of values that are observed in SN~Ibn spectra.
This is because wind velocities of a few thousand km s$^{-1}$ are not infrequent in Wolf-Rayet stars. In addition, spectra of well-monitored SNe~Ibn 
such as SN 2010al \citep{pasto13a} exhibit a FWHM velocity for the He~I lines that grows with time, suggesting an increasing contribution of
ejecta/CSM shocked material to the line flux.}

We also identify narrow, unresolved  H$\alpha$, H$\beta$, [O~II] $\lambda$3727, [O~III] $\lambda$5007, [N~II] $\lambda$6584, and [S~II] $\lambda\lambda$6717, 6731 emissions,
which may originate in the interstellar gas unrelated with the SN, or in the slow-moving, 
unshocked CSM of  LSQ12btw. We favour the former explanation, since the SDSS DR8 spectrum of the host
galaxy\footnote{http://skyserver.sdss3.org/dr8/en/.}  also exhibits these  emission features. This is also supported 
by the facts that the flux ratios between the putative galaxy 
lines remain approximately constant in the pre-SN SDSS spectrum of the host galaxy and in the two spectra of LSQ12btw, 
and that the narrow H$\alpha$ is spatially more extended than the emission attributed to the SN (see Appendix, Figure \ref{figA1}).

\begin{figure}
\includegraphics[scale=.43,angle=0]{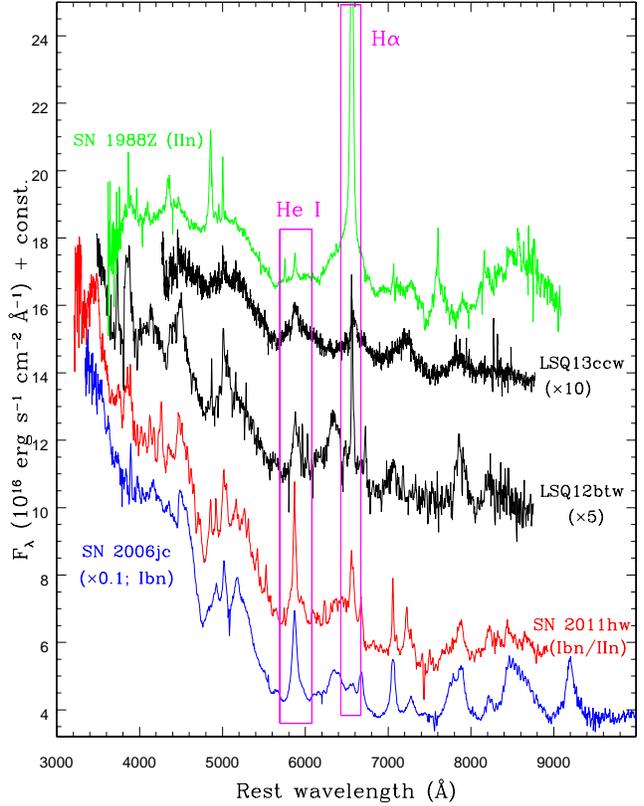}
\caption{Comparison of the Keck-II spectrum of LSQ13ccw (obtained on 2013 September 9) and the NTT spectrum of LSQ12btw (obtained on 2012 April 12),
with spectra of the Type Ibn SN 2006jc \protect\citep{pasto07}, the Type Ibn/IIn SN 2011hw  \protect\citep{pasto13a}, and the prototypical SN~IIn 1988Z \citep{tur93}.
The magenta boxes mark the regions of He I $\lambda$5876 and H$\alpha$ in the 5 spectra.
}
\label{fig6}
\end{figure}

Two spectra of LSQ13ccw were obtained at similar epochs, on 2013 September 9 ($\sim 10.1$ days after the adopted explosion time; this spectrum has poor signal-to-noise) and September 10 ($\sim 11.4$ days).
The spectra are very similar, showing a relatively blue continuum with undulations  at the shorter wavelengths that can be attributed to Fe~II lines. 
At about 5880~\AA\ (rest wavelength) we note a broad feature ($v_{\rm FWHM} \approx 9000$--10,000 km s$^{-1}$) with a narrower P-Cygni profile superimposed, very likely a blend of Na~I~D  and He~I $\lambda$5876.  
This narrow feature indicates the presence of material moving at a relatively low speed ($v \approx 3400$  km s$^{-1}$). 
Additional broad bumps are detected between 6500 and 6700~\AA\ (a blend containing C~II $\lambda\lambda$6578, 6583, He~I $\lambda$6678, and possibly H$\alpha$), at about 7200~\AA\ (a line blend of different species,
probably including [Ca~II] $\lambda\lambda$7291, 7323, C~II $\lambda\lambda$7231, 7236, He~I $\lambda$7065, and He~I $\lambda$7281), and at $\sim 7800$~\AA, observed also in LSQ12btw spectra, and identified as O~I $\lambda$7774. 

The identification of typical He~I lines such as $\lambda$7065 and $\lambda$7281 cannot be firmly confirmed in the two spectra of LSQ13ccw, 
likely because of blending with lines of other ions, and the presence of a broad telluric absorption band which also may affect the profile of the 
observed features. However, the blue pseudo-continuum and the detection of fairly narrow He~I $\lambda$4471, $\lambda$5016, and $\lambda$5876 
lines superposed on broader components suggest that the ejecta are interacting with He-rich CSM, hence supporting the Type Ibn SN classification.
In Figure \ref{fig5} we show the profiles of strong He~I lines and the feature at about 6590~\AA~in the 2013 September 9 Keck-II spectrum of LSQ13ccw, along with those
of the 2012 April 12 NTT spectrum of LSQ12btw. We remark on the presence of relatively narrow P-Cygni absorptions in the He~I features, blueshifted by about 2500 km s$^{-1}$
from the rest wavelengths. 

More controversial is the detection of H$\alpha$ in the spectra of the two SNe. As stated before, in LSQ12btw the unresolved H$\alpha$ is very likely produced by 
background contamination (right panel of Figure \ref{fig5}, and Figure \ref{figA1} in the Appendix).  
In the case of LSQ13ccw, the situation is less clear.
Very narrow H$\alpha$ emission is visible in the 2013 September 10 Keck-II spectrum, but from the two-dimensional spectrum (Figure  \ref{figA1}) it is 
impossible to verify whether this line arises from the SN CSM or from the galaxy background. In addition,
we cannot rule out that H contributes to the intermediate-velocity feature centered at about 6590~\AA\ in the rest-wavelength spectrum, 
although this feature appears to be significantly redshifted (by about 30~\AA) with respect to the H$\alpha$ rest wavelength. In Figure \ref{fig5} (right panel),
the expected position of the core of the likely predominant C~II $\lambda\lambda$6578, 6583 emission is also marked. Accounting for these uncertainties,
we tentatively suggest for LSQ13ccw a Type Ibn/IIn classification, although its putative H$\alpha$ feature is slightly weaker than in other SNe~Ibn/IIn, and 
much weaker than that observed in SNe~IIn \citep[see][and comparison in Figure \ref{fig6}]{pasto08b,smi12,pasto13a}.

\begin{figure*}
\includegraphics[scale=0.54,angle=270]{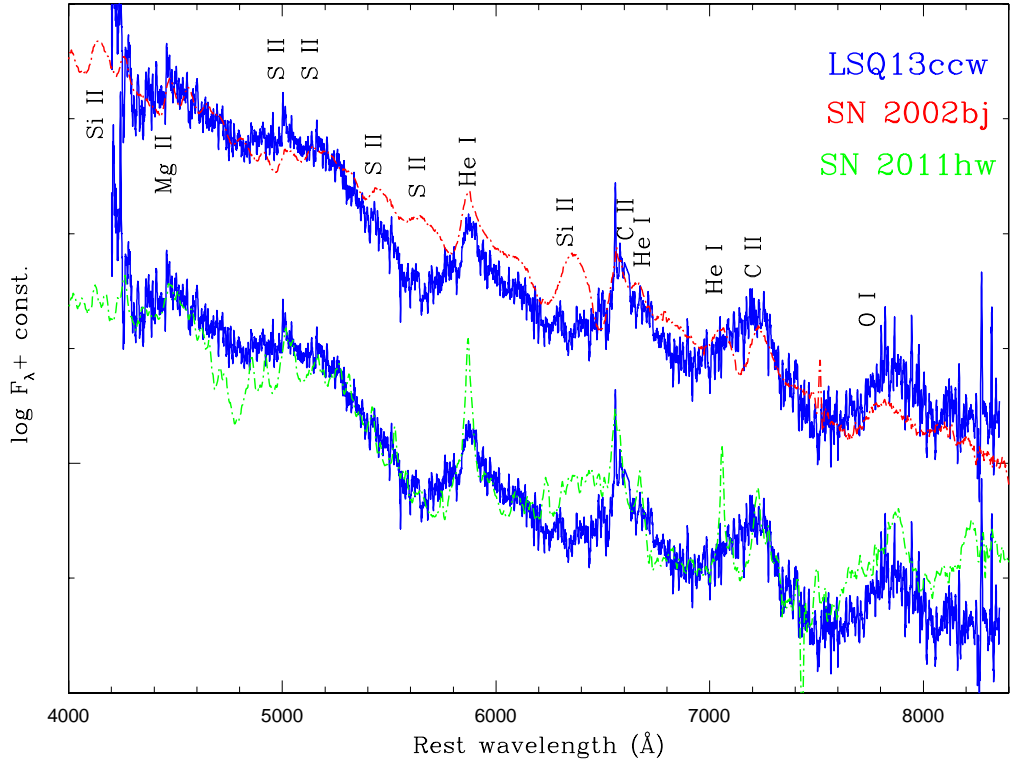}
\caption{ Comparison of the 2013 September 10 Keck-II spectrum of LSQ13ccw (blue solid line) with those of the peculiar SN~Ib 2002bj (red dot-dashed line;
top) 
and the Type Ibn/IIn SN 2011hw (green dot-dashed line; bottom). The most significant lines observed in the spectrum of SN 2002bj \citep{poz10} are also marked.}
\label{fig7}
\end{figure*}

 Some similarity can be also noticed between the spectrum of LSQ13ccw and that of the enigmatic, fast-evolving 
SN~Ib 2002bj \citep[see Fig. \ref{fig7}, and][]{poz10}. The main differences are that the spectrum of SN 2002bj is dominated by lines having 
relatively broad P-Cygni profiles, with C~II, S~II and Si~II features being unequivocally identified (in particular, Si~II $\lambda$6355 is very prominent), 
and it does not show the depression at about 5600 \AA\ typical of SNe~Ibn, which is clearly visible in both LSQ13ccw and SN 2011hw.
In the next section, we will address more comprehensively the implications of the similarity of some SNe~Ibn with SN 2002bj.

In summary, the main spectral properties of the two SNe are those observed in other Type Ibn events (Figs. \ref{fig4}, \ref{fig6}, and \ref{fig7}).
Differences can be noticed, especially in the widths 
of the He~I lines and in the spectral appearance of the region between 6300 and 6700~\AA, that indicate a significant degree of heterogeneity 
in the ejecta/CSM composition. Differences are also observed in the strengths of the O~I and Ca~II features.

\section{Discussion} \label{disc}

We have computed quasi-bolometric light curves, integrating the fluxes in the optical bands at all available epochs. When photometric points in individual bands were missing,
their contribution was estimated by interpolating adjacent photometry. 
The contribution of the missing bands, especially at early epochs, was
extrapolated assuming that the colour evolution of both LSQ12btw and LSQ13ccw were similar to those of other SNe~Ibn observed at early phases \citep[e.g., SNe 2000er and  2010al;\footnote{For these SNe, the synthetic photometry in the Sloan bands has been calculated using the available spectra.}][]{pasto08a,pasto13a}.

\begin{figure}
{\includegraphics[width=3.4in]{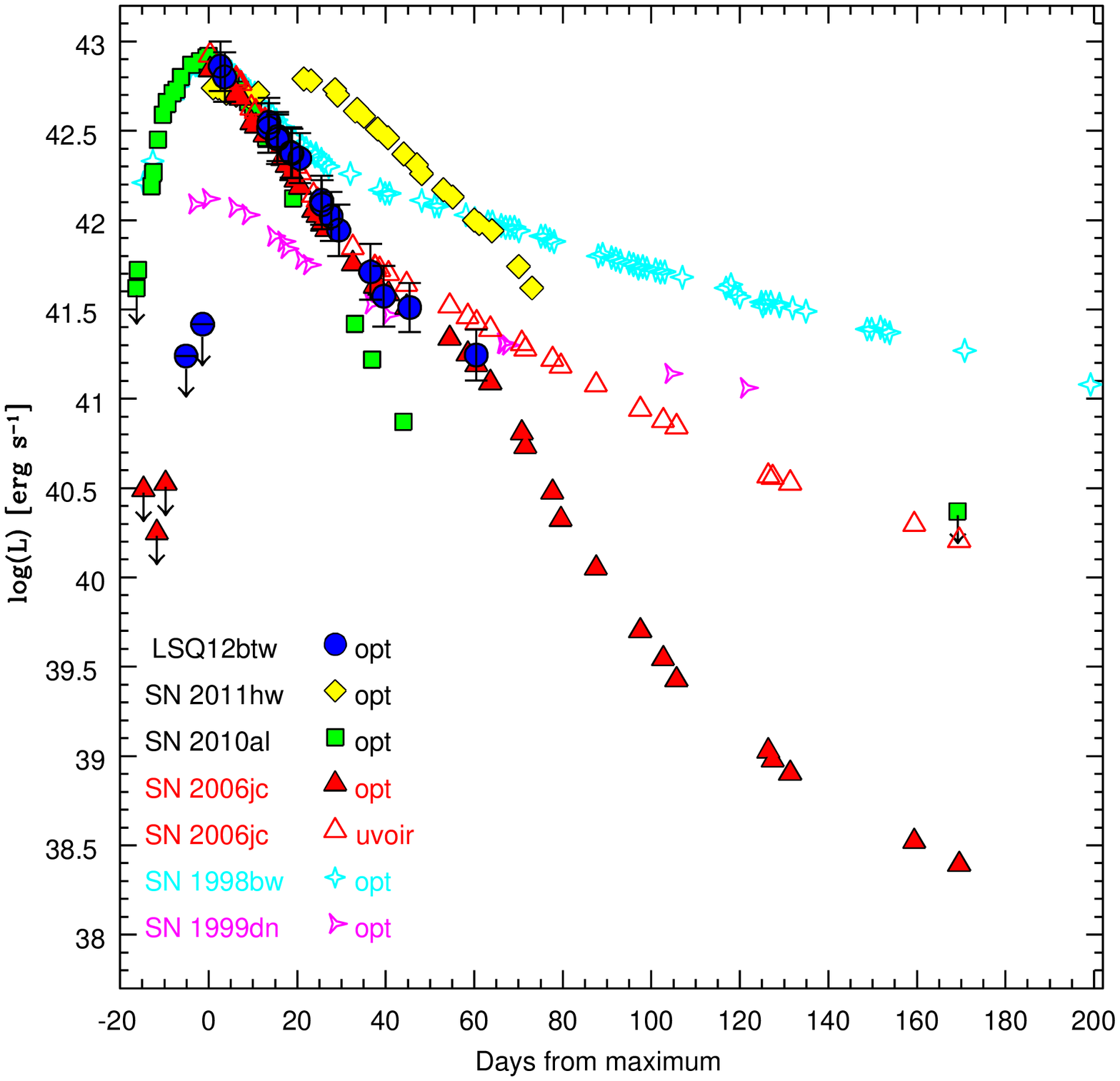}
\includegraphics[width=3.4in]{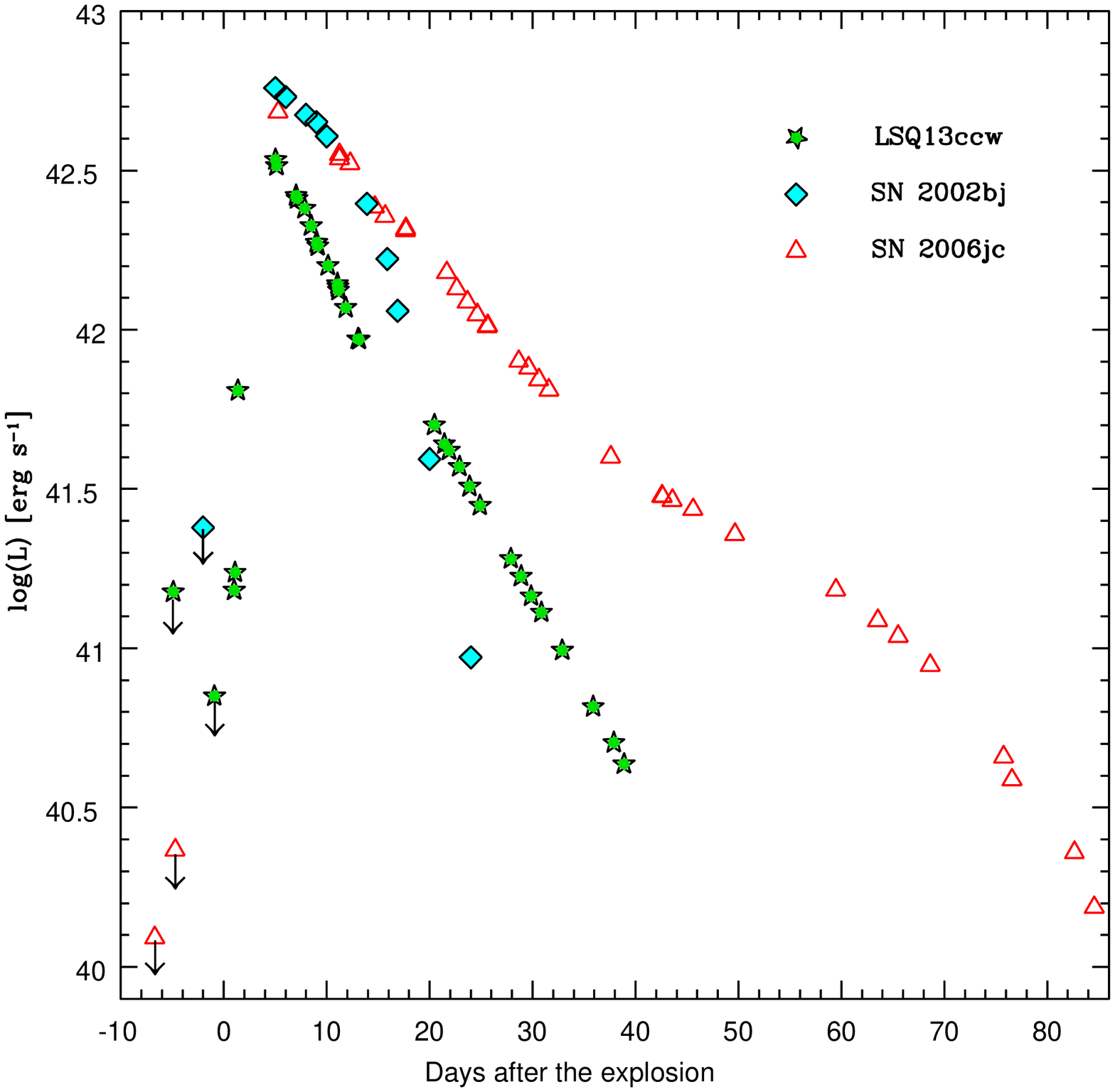}}
\caption{Top: Quasi-bolometric light curve of LSQ12btw, computed integrating the fluxes in the $gVriz$ optical bands 
compared with those of the Type Ibn SNe 2006jc, 2010al, and 2011hw, the Type Ib SN 1999dn, and the Type Ic SN 1998bw (computed integrating the $U$- to $I$-band fluxes; see the text for references).
For SN 2006jc, the ``UVOIR'' light curve obtained from integrating the optical plus NIR fluxes is also shown for comparison. 
Bottom: Quasi-bolometric ($BVgri$ bands) light curve of LSQ13ccw, compared with the curves of SNe 2002bj and 2006jc obtained by integrating the flux contributions of the $BVRI$ bands.  
The stringent prediscovery limits allow us to provide good constraints on the explosion epochs for the three objects.} 
\label{fig8}
\end{figure}

The (optical) quasi-bolometric light curve of  LSQ12btw is shown in Figure \ref{fig8} (top), along with those of
 SNe 2006jc \citep{fol07,pasto07,pasto08a}, 2010al, and 2011hw \citep{pasto13a}, and those of the Type Ib  SN 1999dn \citep{ben11} and 
the luminous Type Ic  SN 1998bw \citep[][and references therein]{pat01}. The peak luminosity of  LSQ12btw, about $6 \times 10^{42}$ erg s$^{-1}$, is high and comparable to those of other SNe~Ibn and the ``hypernova'' SN 1998bw.
Although LSQ12btw is marginally fainter, its light curve is remarkably similar to that of  SN 2006jc. 
In the same Figure~\ref{fig8} (top), we also illustrate the ``UVOIR'' curve (i.e., including the NIR
contribution) of SN 2006jc \citep{pasto07,pasto08a,fol07,imm08,seppo08,smi08,elisa08,anu09}. 
It was shown that in SN 2006jc,  at phases $> 50$ days, much of the flux moved to the NIR domain,
owing to  dust formation in a cool, dense circumstellar shell \citep[e.g.,][]{smi08,seppo08}. 
We do not have NIR observations to confirm  a similar occurrence in LSQ12btw.
Lacking observations at phases later than a few months, we cannot properly constrain the amount of $^{56}$Ni synthesised by  LSQ12btw.
However, based on the striking photometric similarity with SN 2006jc, we argue that the two objects ejected similar amounts of $^{56}$Ni \citep[about $0.3 \pm 0.1$~M$_\odot$; e.g.,][]{pasto08a}.

In Figure \ref{fig8} (bottom) the quasi-bolometric light curve of LSQ13ccw (optical bands only) is compared with the corresponding light curves of SNe 2002bj and 2006jc. Although the evolution of SN 2006jc is fast in the optical bands (top of Fig.~\ref{fig8}), 
one may notice that LSQ13ccw is even faster, declining in luminosity by 2 orders of magnitude
in about 40 days. This rapid evolution is comparable with that experienced by the peculiar, still unique SN~Ib 2002bj.
However, the spectrum of SN 2002bj is somewhat different from that of LSQ13ccw and other SNe~Ibn, since it shows hybrid characteristics between  narrow-lined SNe~Ia and SNe~Ibn.
Prominent P-Cygni lines of He~I were detected in the early-time spectrum of SN 2002bj, together with more typical SN~Ia features, such as Si~II and S~II
\citep{poz10} never unequivocally detected in SN~Ibn spectra. Photometrically it is quite similar to a SN~Ibn, with $M_R \approx -18.5$ mag at 
peak, and it has a similarly fast-declining light curve.
\citet{poz10} suggested that  some of the  properties of SN 2002bj resemble those expected for the so-called .Ia SNe\footnote{These were proposed to be produced in  AM CVn-type binary white dwarfs, where a thermonuclear explosion occurs in the accreted He shell on the 
primary star, producing a fast and faint transient with a He-rich spectrum \citep{bil07}.} \citep{bil07,shen10}, but the arguments offered were not conclusive. Although the .Ia SN scenario is plausible for SN 2002bj,
some similarity shared with SNe~Ibn may suggest further exploration of the core-collapse scenario of massive stars as a viable explanation for this peculiar object.

So far, only 16 objects classified as SNe~Ibn have been discovered, including LSQ12btw and LSQ13ccw (Table \ref{tab4}).
Most of these objects show little or no evidence of hydrogen Balmer lines in their spectra, but do exhibit prevalent He lines of different widths, often with narrow components, and a blue pseudo-continuum likely caused by a blend of broad fluorescent lines of Fe.
This suggests that SNe~Ibn are powered by the interaction of SN ejecta with He-rich CSM, possibly produced during episodic mass-loss events affecting the He-rich layers of the progenitor stars prior to the SN explosion. 
The precursors of SNe~Ibn are likely H-poor WN-type or WCO-type Wolf-Rayet (WR) stars \citep{pasto07,fol07,smi08,tom08}. 
Narrow H lines are clearly visible both in SN 2005la \citep{pasto08b} and SN 2011hw \citep{smi12,pasto13a}, and link these events to massive stars still having a residual H layer.  \citet{smi12} proposed that the precursor of  SN 2011hw was indeed a star that was transitioning 
from a luminous blue variable (LBV) phase to an early WR phase. 
However, we argue that, since the H lines observed in the spectra of  LSQ12btw and LSQ13ccw are weak or unrelated to the most recent progenitor evolution (cf. Section \ref{s_sp}), their progenitors were 
{\bf H depleted} at the time of explosion, and these events match a more canonical WN or WCO progenitor scenario.

Observations show  a significant degree of heterogeneity within the Type Ibn SN family. In fact, the SN~Ibn SN zoo
includes objects with fast-evolving light curves (e.g., LSQ13ccw), and objects with a relatively broad light curve resembling those of normal stripped-envelope SNe \citep[e.g., SN 2010al;][]{pasto13a}. 
Some others even have double- or multi-peaked light curves 
\citep[SNe 2005la, 2011hw, and iPTF13beo;][]{pasto08b,smi12,pasto13a,gor13}, and at least one object having a very slow late-time luminosity evolution \citep[OGLE-2012-SN-006;][]{pasto13b}.
This variety is probably related to intrinsic differences in the final properties of the progenitors and/or the configuration and composition of their CSM. 
The existence of transitional SNe~Ibn showing evidence of circumstellar H (SN 2005la and 2011hw) suggests a continuity
between stripped-envelope SNe and some SNe~IIn \citep{tura13}. 
The diversity is also illustrated with the discovery of the first stripped-envelope event, SN 2010mb \citep{ben14}, that shows evidence of interaction with H-free CSM. Its spectra show  narrow lines of $\alpha$-elements (in particular, the collisionally excited [O~I] $\lambda$5577 line) instead of the more canonical narrow H and He~I lines.\footnote{According to the SN classification criteria, an object with these characteristics should be formally designated a SN~Icn.} 
This indicates that even  He-free WR stars may experience significant mass-loss events a very short time before the terminal SN explosion.

\subsection{Supernovae Ibn and their host environment}

\begin{table*}            
\tiny
\caption{Basic information on the host galaxies of our SN sample. 
}
\centering
\begin{tabular}{lcccccccccc}
\hline\hline
SN                  & Galaxy              & Type  &  z &  $\mu$ & $B_{\rm tot}$ & $M_{\rm B,tot}$ & R$_{\rm 0,SN}$/R$_{\rm 25}$ & 12+log(O/H) & 12+log(O/H)$_{\rm SN}$ & Reference\\
                       &                          &    &         &   (mag)   & (mag)     & (mag)          &                             &            (dex)       &   (dex)            &          \\
\hline
1999cq               &   UGC 11268              & Sbc     & 0.0271$^\P$  & 35.27 & 14.53        & $-$21.52       & 0.13           &  9.24 &  9.18 & A \\           
2000er               &   PGC 9132               & Sab     & 0.0302$^\P$  & 35.52 & 14.81        & $-$21.68       & 0.52$^\diamond$ &  9.27 &  9.06 & B \\           
2002ao               &   UGC 9299               & SABc    & 0.0054$^\P$  & 31.73 & 14.57        & $-$18.07       & 0.60           &  8.60 &  8.36 & B,C \\        
2005la               &   KUG 1249+278           & Sc      & 0.0190$^\P$  & 34.49 & 16.67        & $-$18.32       & 0.94           &  8.64 &  8.27 & D \\           
2006jc               &   UGC 4904               & SBbc    & 0.0061$^\P$  & 32.01 & 15.05        & $-$17.29       & 0.70           &  8.45 &  8.17 & B,C,E \\     
2010al               &   UGC 4286               & Sab     & 0.0172$^\P$  & 34.27 & 14.58        & $-$20.44       & 0.37           &  9.04 &  8.89 & F \\           
2011hw               &   anonymous              & SBbc    & 0.023        & 34.92 & 16.02        & $-$19.47       & 0.64           &  8.86 &  8.60 & F,G \\        
PS1-12sk             &   CGCG 208-042           & E       & 0.0546$^\P$  & 36.84 & 15.50        & $-$21.71       &  --            &   --  &   --  & H \\           
OGLE-2012-006$^\dag$  & anonymous                & Sd/Irr  & 0.057   & 36.94 & 21.33            & $-$15.92       & 0.93           &  8.20 &  7.83 & I,J \\        
iPTF13beo            & SDSS J161226.53+141917.7 & Sp-ab   & 0.091   & 38.01 & 20.23$^\ddag$     & $-$18.03       & 0.55           &  8.59 &  8.37 & K \\           
LSQ12btw             & SDSS J101028.69+053212.5 & Sab     & 0.0576  & 36.97 & 18.08$^\ddag$     & $-$19.33       & 0.50           &  8.83 &  8.63 & L \\           
LSQ13ccw             & PGC 866755               & Sbc     & 0.0603  & 37.07 & 16.08            & $-$21.29       & 0.92           &  9.19 &   8.83 & L \\           
CSS140421$^\dag$      & SDSS J142041.70+031603.6 &  --     & 0.07        & 37.41 & 20.46$^\ddag$ & $-$17.27$^\S$  &  --            &  8.45 &   --  &  M \\          
2014av               & UGC 4713                 & Sb      & 0.0308$^\P$  & 35.56 & 14.09        & $-$21.82       & 0.25           &  9.29 &   9.19 &   N   \\      
2014bk               & SDSS J135402.41+200024.0 & S       & 0.0697       & 37.40 & 19.52        & $-$18.16       &  --            &  8.61 & -- & O \\              
ASASSN-14dd          & NGC 2466                 & Sc      & 0.0168$^\P$  & 34.22 & 13.72        & $-$21.12       & 0.76           &  9.16 & 8.86 & P,Q \\  \hline 
\end{tabular}

\begin{flushleft}

Note: After the SN and host-galaxy designations (columns~1 and 2), we report  the galaxy type (column~3), the redshift (from the Virgo-infall corrected  recession velocity, when available; column~4), the
distance modulus (assuming H$_0$ = 73 km s$^{-1}$ Mpc$^{-1}$; column~5),
the total $B$-band magnitude of the host galaxy (column~6), 
the absolute total $B$-band magnitude of the host galaxy corrected for Galactic \citep{sch11} and internal extinction 
and $K$-corrected following the HyperLeda prescriptions (column~7), 
the $R_{\rm 0,SN}/R_{\rm 25}$ ratio (see text, column~8), the integrated oxygen abundance of the host galaxy 
following \protect\citet[][reported to the Tremonti H$_0$ = 70 km s$^{-1}$ Mpc$^{-1}$ scale, to simplify comparison; column~9]{tre04}, the oxygen abundance at the SN position assuming the average radial dependence as in \citet[][]{pil04} (column~10), 
and the main references for the SN (column~11).
Sources of information include HyperLeda, the NASA/IPAC Extragalactic Database (NED), and, when information was not available, our analysis of original frames (column~11).

A = \protect\cite{mat00};
B = \protect\cite{pasto08a};
C = \protect\cite{fol07};
D = \protect\cite{pasto08b}; 
E = \protect\cite{pasto07};
F = \protect\cite{pasto13a};
G = \protect\cite{smi12};
H = \protect\cite{san13};
I = \protect\cite{pri13};
J = \protect\cite{pasto13b};
K = \protect\cite{gor13};
L = this paper;  
M = \protect\cite{polsh14};
N = \protect\cite{xu14};
O = \protect\cite{mor14};  
P = \protect\citep{sta14};
Q = \protect\citep{pri14}. \\
$^\diamond$Computed assuming $R_{\rm 0,SN}$ equal to projected radius. \\
$^\dag$OGLE-2012-006 = OGLE-2012-SN-006 and CSS140421 = CSS140421:142042+031602. \\
$^\ddag$Converted from Sloan magnitudes adopting the conversion relations from \texttt{Lupton, 2005 (https://www.sdss3.org/dr8/algorithms/sdssUBVRITransform.php\#Lupton2005).} \\
$^\S$Adopting an average internal absorption $A_i=0.2$ mag.
$^\P$Redshift computed from $v_{\rm Vir}$ (source Hyperleda). 
\end{flushleft}
\label{tab4}
\end{table*}

Thus far, it has never been properly investigated whether there is a connection between the occurrence of rare SNe~Ibn, and chemical or morphological properties of their parent galaxies. 
Table \ref{tab4} presents basic information on the host-galaxy properties for all 16 SNe~Ibn, and we have homogenised the available (and sometimes sparse) data to common systems.
The primary source of the morphological types was HyperLeda, and in a few cases NED or SDSS. In six cases we relied on the deepest images collected during the follow-up campaigns carried 
out by our team (column~11). Distance moduli were obtained from the Virgo-corrected recessional velocities, assuming H$_0$ = 73 km s$^{-1}$ Mpc$^{-1}$.
In four cases\footnote{The galaxy parameters were determined for SNe 2011hw, LSQ13btw, OGLE-2012-SN-006 and iPTF13beo, CSS140421:142042+031602} we performed detailed photometric analyses to determine $B_{\rm tot}$, inclinations, and position angles required to derive the deprojected SN relative distances from the nuclei, 
$R_{\rm 0,SN}/R_{\rm 25}$,\footnote{$R_{\rm 25}$ is the isophotal radius for the $B$-band surface brightness of 25 mag arcsec$^{-2}$.} determined following \citet{hak09}. 
The total absolute magnitudes were estimated using the Galactic extinctions taken from \citet{sch11}, and the internal absorptions from HyperLeda or computed with the same prescriptions.
The average oxygen abundance estimates of the hosts (column~9) were computed according to the luminosity-metallicity relation of \citet{tre04}.
With the aim of disentangling possible effects due to radial gradients, we have also computed the metallicity at the SN radial distance. 
We assumed that Tremonti's metallicity is dominated by the contribution of the innermost regions of the host galaxy, and applied the average radial dependence of 
the oxygen abundance computed for the nearby sample of spiral galaxies of \citet{pil04}. The oxygen abundances at the distances of the SNe are reported in column 10 of Table \ref{tab4}.

Since none of the host galaxies are closer than the Virgo cluster, none have been studied in great detail.
We note that among the 37 stripped-envelope CCSNe (SESNe, including SNe~Ib/c and IIb) discovered during the decade 1999--2008 and closer than 30~Mpc (data from the Asiago Supernova Catalog), only 
SNe 2002ao and 2006jc exhibited easily recognisable He~I emission features indicative of interaction with He-rich CSM. Therefore, it seems reasonable to assume that the relative discovery rate of SN~Ibn/SESNe 
is $\sim 5.4$ per cent, implying a relative ratio SN~Ibn/CCSNe $\approx 2$ per cent.

With a single notable exception, the galaxies hosting SNe~Ibn are spirals, suggesting that these SNe are associated with young stellar populations. 
However, the discovery of the Type Ibn SN~PS1-12sk \citep{san13} poses new questions on the massive-star scenario for this class of objects.
PS1-12sk exploded in the galaxy cluster RXC J0844.9+4258, in the outskirts of the luminous elliptical galaxy CGCG 208-042.
The detailed analysis presented by \citet{san13} on the possible  scenario involving degenerate progenitors still leaves some unclarified issues
(e.g., presence of He, H, dense CSM, ejected $^{56}$Ni mass, etc.). 
Though there is no direct evidence of star formation at the site of the explosion, the activity of the nucleus and  the indications for the presence of cooling flows suggest that PS1-12sk is one of the rare (but known) 
cases of CCSNe exploding in early-type galaxies with some residual star-formation activity \citep{hakobyan08}.   
On the other hand, the major outburst observed in SN 2006jc two years before the SN explosion can be comfortably explained with instabilities in massive stars during the final stages of their life \citep{smiarn14},
while it is hard to explain in a picture where the progenitor is a white dwarf.

The intrinsic luminosities of the host galaxies span a range of over 6 magnitudes, from the dwarf hosts of  OGLE-2012-SN-006 and SN 2006jc ($M_B \approx -15.8$ to $-17.3$ mag) to the bright spirals hosting SNe 2000er and 2014av ($M_B \leq -21.5$ mag). 
Such a wide luminosity range is reflected in a large spread of the integrated O abundances of the hosts  \citep[via the luminosity-metallicity relation of ][]{tre04} and the explosion sites (1.3 dex, column~10 of Table~\ref{tab4}). 
The available information on the explosion sites, though scanty and based on statistical relations, thus seems to exclude the possibility that metallicity plays a significant role in determining the late-time evolution of the WR progenitors and their explosion as SNe~Ibn.

In conclusion, a growing wealth of evidence indicates that the observed properties of SNe~Ibn may be consistent with the explosion of massive WR \citep[e.g.,][]{fol07,pasto07,tom08} or transitional LBV/WR precursors \citep{smi12,pasto13a}
that recently enriched their CSM  with He-rich material through major eruptive events occurring shortly before the stellar core collapse, although we cannot rule out that binarity may
play a significant role in the pre-SN evolution of the progenitor star.
The sequence of events preceding core collapse is not fully understood and requires some fine tuning.
We eagerly await the explosion of a very nearby SN~Ibn, with direct information on the progenitor system, that will help solve the remaining open issues on the nature of this rare SN subclass.

\section*{Acknowledgments}

This work is partially based on observations obtained under the ESO-NTT programs with IDs 184.D-1140 and ID 188.D-3003,
with the latter being part of PESSTO (the Public ESO Spectroscopic Survey for Transient Objects Survey).
We are grateful to S. Bradley Cenko, Kelsey I. Clubb, and WeiKang Zheng for their help with observations of LSQ13ccw at the Keck-II telescope,
and to M. L. Pumo and S. Schulze for useful discussions.

A.P., E.C., S.B., and M.T. are partially supported by the PRIN-INAF 2011 with the project {\it ``Transient Universe: from ESO Large to PESSTO.''}
Research leading to these results has received funding from the European Research Council under the European
Union's Seventh Framework Programme (FP7/2007-2013)/ERC Grant agreement No. [291222] (PI: S.J.S.) and 
EU/FP7-ERC grant No. [307260] (PI: A.G.-Y.). S.J.S. is also supported by STFC grants ST/I001123/1 and ST/L000709/1.
A.G.-Y. is also supported by The Quantum Universe I-Core program by the Israeli Committee for planning
and funding, the ISF and GIF grants, and the Kimmel award. This research used resources of the National 
Energy Research Scientific Computing Center, which is supported by the Office of Science of the U.S. Department of Energy 
under Contract No. DE-AC02-05CH11231. This work was partly supported by the European Union FP7 programme through 
ERC grant number 320360. N.E.R. acknowledges support from the European Union Seventh Framework Programme (FP7/2007-2013) 
under grant agreement No. 267251 ``Astronomy Fellowships in Italy'' (AstroFIt).
A.M.G. acknowledges financial support by the MICINN grant AYA2011-24704/ESP, by the ESF EUROCORES Program EuroGENESIS 
(MINECO grants EUI2009-04170), SGR grants of the Generalitat de Catalunya, and by the EU-FEDER funds. 
M.D.S. gratefully acknowledges generous support provided by the Danish Agency for Science and Technology and Innovation realised
through a Sapere Aude Level 2 grant. S.T. acknowledges support by the Transregional Collaborative Research Centre 
TRR 33 {\it ``The Dark Universe''} of the DFG.
A.V.F.'s supernova group at UC Berkeley received support through NSF grant
 AST--1211916, the TABASGO Foundation, Gary and Cynthia Bengier, the 
Richard and Rhoda Goldman Fund, and the 
Christopher R. Redlich Fund.

This paper is based on observations made with the Italian Telescopio Nazionale Galileo
(TNG) operated on the island of La Palma by the Fundaci\'on Galileo Galilei of
the INAF (Istituto Nazionale di Astrofisica). It is also based on observations made with the Liverpool
Telescope (LT) and the Gran Telescopio Canarias (GTC), operated on the island of La Palma at the Spanish 
Observatorio del Roque de los Muchachos of the Instituto de Astrofisica de Canarias. 
Some of the data presented herein were obtained at the W. M. Keck Observatory, which is operated as a scientific partnership among 
the California Institute of Technology, the University of California, and the National Aeronautics and Space Administration (NASA). 
The Observatory was made possible by the generous financial support of the W. M. Keck Foundation.
 This work makes use of observations from the LCOGT network, and utilises data from the 40-inch ESO Schmidt Telescope at the La Silla Observatory in Chile  with the large-area QUEST 
camera built at Yale University and Indiana University.

The Pan-STARRS1 (PS1) Surveys have been made possible through contributions of the Institute for Astronomy, the University of Hawaii, the Pan-STARRS Project Office, the Max-Planck Society and its participating institutes, the Max Planck Institute for Astronomy, Heidelberg and the Max Planck Institute for Extraterrestrial Physics, Garching, The Johns Hopkins University, Durham University, the University of Edinburgh, Queen’s University Belfast, the Harvard-Smithsonian Center for Astrophysics, the Las Cumbres Observatory Global Telescope Network, Inc., the National Central University of Taiwan, the Space Telescope Science Institute, NASA under Grant No. NNX08AR22G issued through the Planetary Science Division of the NASA Science Mission Directorate, the US NSF
under grant AST--1238877, and the University of Maryland.

This research has made use of the NASA/IPAC Extragalactic Database (NED) which is operated by the Jet Propulsion 
Laboratory, California Institute of Technology, under contract with NASA.
We acknowledge the usage of the HyperLeda database (http://leda.univ-lyon1.fr).


\appendix
\section[]{Narrow lines in the spectra of LSQ12btw and LSQ13ccw} \label{app}

\begin{figure*}
\includegraphics[scale=.51,angle=0]{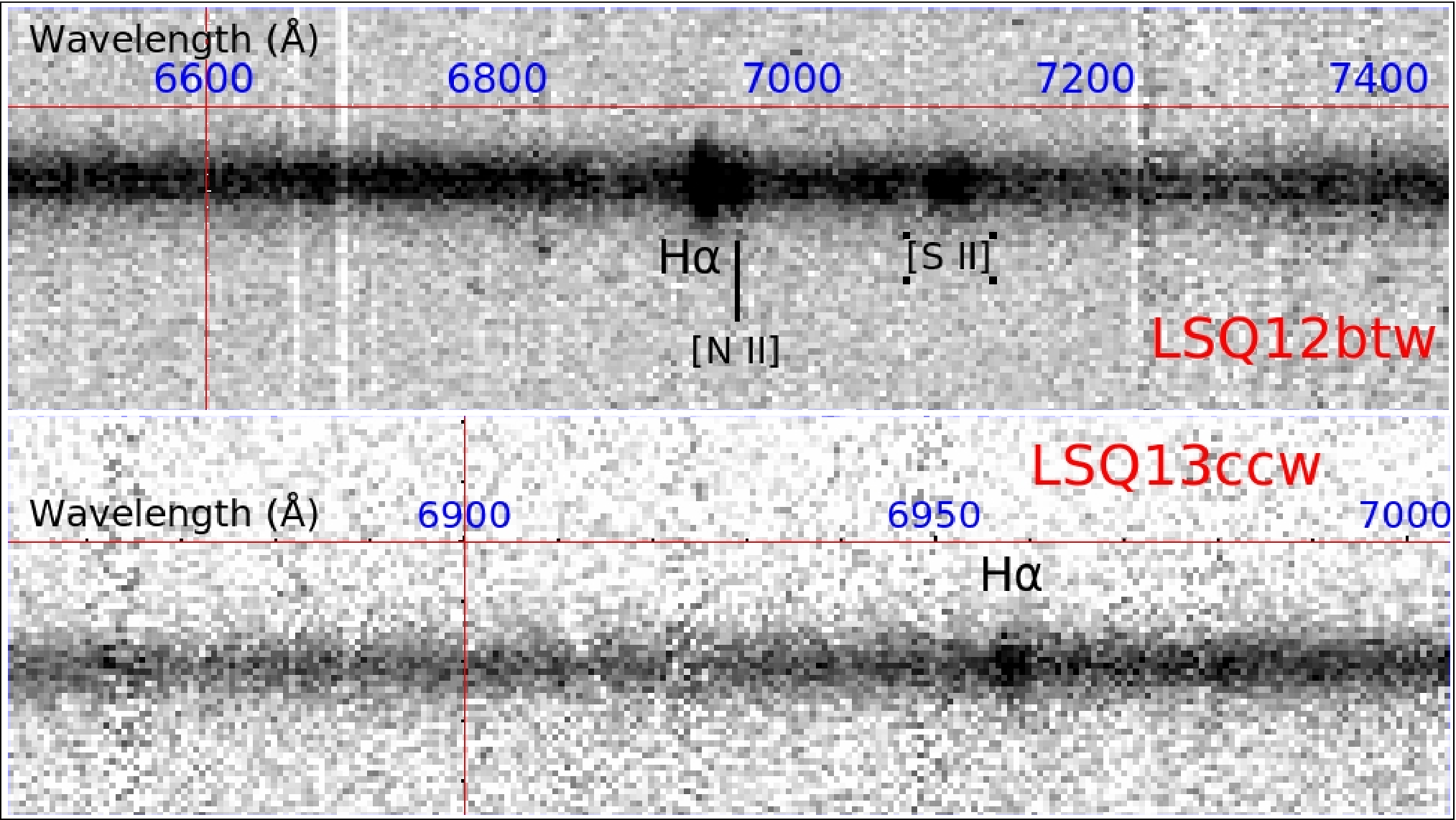}
\caption{Zoom-in on the H$\alpha$ region in the highest-resolution two-dimensional
spectra of the two SNe: NTT spectrum of LSQ12btw taken on 2012 August 21 (top) and  
Keck-II spectrum of LSQ13ccw obtained on 2013 September 9 (bottom). Both spectra are shown at the host-galaxy frame.} 
\label{figA1}
\end{figure*}

In Figure \ref{figA1}, we show a close-up view of the H$\alpha$ region in our highest-resolution, two-dimensional spectra of LSQ12btw and LSQ13ccw.
In the spectrum of LSQ12btw, narrow H$\alpha$, [N~II], and [S~II] are detected, with a spatial extension which is broader than the SN emission. For this reason, we
believe that these lines are physically unrelated to the SN. In the spectrum of LSQ13ccw, the narrow H$\alpha$ emission does not show spatially extended wings broader than
the continuum. Therefore, in this case we cannot firmly establish or rule our the association of this feature with the SN. 

\label{lastpage}

\end{document}